\documentclass{emulateapj}           %apj-style

\usepackage{lscape}

\newcommand{\Spitzer}{ {\it Spitzer} }

\newcommand{\Oii}{[O~{\tiny II}] $\lambda$3727}
\newcommand{\Oiii}{[O~{\tiny III}] $\lambda$5007}
\newcommand{\OiiiD}{[O~{\tiny III}] $\lambda\lambda$5007,~4959}
\newcommand{\Civ}{C~{\tiny IV} $\lambda$1549}
\newcommand{\Ciii}{C~{\tiny III}] $\lambda$1909}
\newcommand{\Mgii}{Mg~{\tiny II} $\lambda$2798}
\newcommand{\Ha}{H$\alpha$}
\newcommand{\Hb}{H$\beta$}
\newcommand{\Sii}{[S~{\tiny II}] $\lambda$6725}
\newcommand{\Nev}{[Ne~{\tiny V}] $\lambda$3426}
\newcommand{\Cahk}{CaHK $\lambda\lambda$3933,~3969}
\shorttitle{A LARGE FRACTION OF AGN WITH HIGH \emph{F}(24 $\mu$m)/\emph{F}(\emph{R})
RATIO}
\shortauthors{SACCHI ET AL.}

\submitted{Accepted 2009 July 17; Astrophysical Journal in press}

\begin{document}

\title{Spectroscopic identifications of \emph{Spitzer} sources in the SWIRE/\emph{XMM-NEWTON}/ELAIS-S1 field:
       a large fraction of Active Galactic Nuclei with high \emph{F}(24 \micron)/\emph{F}(\emph{R}) ratio}

   \author{N. Sacchi\altaffilmark{1,2},
          F.  La Franca\altaffilmark{1},
          C. Feruglio\altaffilmark{3,4},
          F. Fiore\altaffilmark{3},
          S. Puccetti\altaffilmark{5},
          F.  Cocchia\altaffilmark{3},
          S. Berta\altaffilmark{6},
          M. Brusa\altaffilmark{6},
          A. Cimatti\altaffilmark{7},
          A. Comastri\altaffilmark{8},
          A. Franceschini\altaffilmark{9},
          C. Gruppioni\altaffilmark{8},
          R. Maiolino\altaffilmark{3},
          I. Matute\altaffilmark{1,10},
          M. Polletta\altaffilmark{11},
          L. Pozzetti\altaffilmark{8},
          F. Pozzi\altaffilmark{7},
          C. Vignali\altaffilmark{7},
          G. Zamorani\altaffilmark{8},
          S. Oliver\altaffilmark{12},
          M. Rowan-Robinson\altaffilmark{13},
          G. Smith\altaffilmark{14},
          C. Lonsdale\altaffilmark{15}}

\altaffiltext{1}{ Dipartimento di Fisica, Universit\`a Roma Tre,
       via della Vasca Navale 84, I-00146 Roma, Italy;
email: {\tt sacchi@fis.uniroma3.it, lafranca@fis.uniroma3.it}}
\altaffiltext{2}{TNG Fundaci\'on Galileo Galilei - INAF,
       Rambla Jos\'e Ana Fern\'andez P\'erez 7, 38712 Bre\~na Baja, TF - Spain}
\altaffiltext{3}{INAF-Osservatorio Astronomico di Roma,
       via Frascati 33, I-00040, Monteporzio Catone, Italy}
\altaffiltext{4}{CEA, Irfu, Service d'Astrophysique, Centre de Saclay, F-91191 Gif-sur-Yvette, France}
\altaffiltext{5}{ASI Science Data Center, via G. Galilei, I-00044 Frascati, Italy}
\altaffiltext{6}{Max Planck Institut f\"ur Extraterrestrische Physik, Giessenbachstrasse
1, D-85748 Garching, Germany }
\altaffiltext{7}{Dipartimento di Astronomia, Universit\`a di Bologna,  via Ranzani 1, I-40127 Bologna, Italy }
\altaffiltext{8}{INAF-Osservatorio Astronomico di Bologna,  via Ranzani 1, I-40127 Bologna, Italy }
\altaffiltext{9}{Dipartimento di Astronomia, Universit\`a di  Padova,  vicolo dell'Osservatorio 2, I-35122 Padova, Italy }
\altaffiltext{10}{INAF-Osservatorio Astronomico di Arcetri,  largo E. Fermi 5, I-50125 Firenze, Italy }
\altaffiltext{11}{INAF-IASF Milano,  via E. Bassini 15, I-20133 Milano, Italy }
\altaffiltext{12}{Astronomy Centre, Department of Physics and Astronomy, University of Sussex, Falmer, Brighton BN1 9QJ, UK}
\altaffiltext{13}{Astrophysics Group, Blackett Laboratory, Imperial College of Science Technology
and Medicine, Prince Consort Road, London SW7 2BZ, UK}
\altaffiltext{14}{Center for Astrophysics and Space Sciences, University of California, San Diego, La Jolla, CA 92093-0424, USA}
\altaffiltext{15}{Infrared Processing and Analysis Center, California Institute of Technology 100-22,
Pasadena, CA 91125, USA }

\begin{abstract}

We present a catalog of optical spectroscopic identifications of
sources detected by {\Spitzer} at 3.6 or 24 $\mu$m down to
$\sim$10 and $\sim$280 $\mu$Jy, respectively, in the
SWIRE/\emph{XMM-Newton}/ELAIS-S1 field and classified via line width analysis
and diagnostic diagrams. A total of 1376 sources down to
$R\sim24.2$ mag have been identified (1362 detected at 3.6 $\mu$m,
419 at 24 $\mu$m, and 405 at both) by low-resolution optical
spectroscopy carried out with FORS2, VIMOS, and EFOSC2 at the Very Large Telescope
and 3.6m ESO telescopes. The spectroscopic campaigns have been
carried out over the central 0.6 deg$^2$ area of ELAIS-S1 which,
in particular, has also been observed by \emph{XMM-Newton}  and
{\it Chandra}. We find the first direct optical spectroscopic
evidence that the fraction of active galactic nuclei (AGN; mostly AGN2) increases with
increasing \emph{F}(24 $\mu$m)/\emph{F}(\emph{R}) ratio, reaching values of
70($\pm$20)\%  in the range 316$<$\emph{F}(24 $\mu$m)/\emph{F}(\emph{R})$<$1000. We
present an Infrared Array Camera--Multiband Imaging Photometer color--color diagram able to separate AGN1
from obscured AGN2 candidates.
 After having corrected for the spectroscopic incompleteness of our sample, it results that the AGN fraction at \emph{F}(24 $\mu$m)$\sim$0.8 mJy is $\sim$22($\pm$7)\%  and decreases slowly to $\sim$19($\pm$5)\% down to \emph{F}(24 $\mu$m)$\sim$0.3 mJy.
 \end{abstract}


%%%%%%%%%%%%%%%%%%%%%%%%%%%%%%%%%%%%%%%%%%%%%%%
\section{Introduction}

Multi-wavelength surveys are fundamental instruments to study the cosmological evolution
of extragalactic sources. Each region of the electromagnetic spectrum provides information
on different physical mechanisms which take place in the galaxies.  The emission in the mid-infrared (MIR) band
is mainly produced by dust heated either by the stars (particularly during periods of strong star formation activity)
or by the emission originating in an active galactic nucleus (AGN). In several cases, both causes contribute to the
formation of the total spectral energy distribution (SED) in the MIR (see, e.g., Polletta et al. 2008; Donley et al. 2008 and references therein).

In recent years, \emph{XMM-Newton}- and \emph{Chandra}-based X-ray surveys have been able to
delineate the density and evolution of X-ray unabsorbed and even moderately absorbed AGN (Compton-thin; N$_H$=10$^{22}$-10$^{24}$ cm$^{-2}$)
up to redshift $\sim$4 (see, e.g., Ueda et al. 2003; La Franca et al.  2005).
The fraction of absorbed AGN is observed to decrease with increasing luminosity (e.g., Ueda et al. 2003;
La Franca et al. 2005; Hasinger 2008; see also Lawrence \& Elvis 1982) and to
increase with increasing redshift (La Franca et al. 2005; Ballantyne et al. 2006; Treister \& Urry 2006; Hasinger 2008; Treister et al. 2009).
In this framework, in order to reproduce the cosmic X-ray background, a population of heavily
X-ray absorbed AGN (Compton-thick; CT) as numerous as the Compton-thin one is expected.
Because of the strong X-ray absorption, these CT AGNs are difficult to select even in the X-ray band.

As the AGN (absorbed) activity is expected to contribute significantly to the MIR SED,
observations in the MIR regime are potentially useful for selecting AGN, including those
which are undetected in the X-ray band.

The launch of the {\it Spitzer} telescope has allowed the construction of several
large area multi-wavelength databases with MIR coverage, such as (for example) the
9 deg$^2$ NOAO Deep Wide-Field Survey (NDWFS) of the Bo\"otes field
(Murray et al. 2005), the 3.7 deg$^2$ Extragalactic
First Look Survey (E-FLS; Fadda et al. 2006) and the smaller 0.5 deg$^2$ of the  All-Wavelength Extended Groth Strip International Survey (AEGIS;
Davis et al. 2007).
Indeed, the use of  {\it Spitzer} colors has been revealed to be efficient (if not complete) in selecting
AGN (e.g., Lacy et al. 2004, 2007; Stern et al. 2005).  Using both {\it Spitzer} and X-ray observations,
Brand et al. (2006), Treister et al. (2006), and Donley et al.  (2008) have evaluated the fraction of AGN as a function of the
24 $\mu$m flux, which results in being around 30\%-45\% at $\sim$3 mJy, decreasing to about 10\% at $\sim$0.4 mJy.

Many studies have shown evidence of the existence of X-ray absorbed (Compton-thick) \emph{Spitzer}-selected  AGN
(e.g., Donley et al. 2005, 2007; Alonso-Herrero et al. 2006; Polletta et al. 2006; Steffen et al. 2007;
Alexander et al. 2008). Maybe the most efficient criterion
in selecting X-ray-absorbed AGN is based on selecting
sources with large \emph{F}(24 $\mu$m)/\emph{F}(\emph{R}) ratios, in some cases linked to radio
observations (e.g.,  Mart\'inez-Sansigre et al. 2005, 2007).  Because of the
faint optical fluxes these sources are difficult to spectroscopically identify
in the optical, and, in fact, the AGN classifications for this class of objects are mainly based on
MIR spectroscopy with the Infrared Spectrograph (IRS) on board {\it Spitzer} (e.g., Houck et al. 2005;
Weedman et al. 2006;  Yan et al 2007; Brand et al. 2008). Because of the low resolution of the
IRS spectra, for many sources the classification and the redshift measure are derived from  the identification of the
9.7 $\mu$m silicate absorption features on approximate  power law spectral shapes.

Using stacked {\it Chandra} images
of sources with large \emph{F}(24 $\mu$m)/\emph{F}(\emph{R}) ratios, Daddi et al. (2007) and Fiore et al. (2008,  2009)
have shown that the {\em average} X-ray spectrum can be reproduced
if a high percentage  ($\sim$80\%) of the sources in the sample
are highly X-ray absorbed (even Compton-thick) AGN (but see Donley et al. (2008) and
Pope et al. (2008) for partly different conclusions).
This result is itself important for the understanding
of AGN density evolution, but it needs to be complemented with a
quantification of the real fraction of AGN
among these sources. In fact, the above studies are mainly based on photometric redshifts and
lack direct optical spectroscopic identifications and classifications.
Such measurements are crucial in order
to understand how much of the average resulting hard X-ray spectrum is diluted and/or contaminated by
starburst galaxies  included in the samples (see, e.g., Polletta et al. 2008; Donley et al. 2008).

Brand et al. (2007) have studied in the near IR the spectrum of 10
ULIRG sources with 24  $\mu$m fluxes larger than 0.8 mJy and
large \emph{F}(24 $\mu$m)/\emph{F}(\emph{R}) ratios and found that the SEDs were
compatible with a mixed contribution of AGN and starburst
activity. Studying the SED and MIR spectrum of 21 obscured AGNs
with large \emph{F}(24 $\mu$m)/\emph{F}(\emph{R}) ratios, and  24  $\mu$m fluxes
larger than 1 mJy, Polletta et al. (2008)  found that the
contribution by the starburst component to the bolometric
luminosity was below 20\%.

Dey et al. (2008) have measured spectroscopic redshift for 86 very
luminous dust-obscured galaxies with \emph{F}(24 $\mu$m)$>$0.3 mJy and
\emph{F}(24 $\mu$m)/\emph{F}(\emph{R})$>$1000 in the Bo\"otes field, and find a broad
redshift distribution centered at $z\sim2$. Roughly half the
redshifts are the results of {\em Spitzer} IRS observations, and
half come from ground-based optical or NIR spectroscopy.
However, no optical spectral classification of the galaxies has
been discussed in their work.

Here we present the optical spectroscopic identifications and classification of 1376 sources of the
central 0.6 deg$^2$ of the ELAIS-S1 field included in the {\it Spitzer} Wide-area Infrared
Extragalactic Survey (SWIRE; Lonsdale et al. 2003, 2004) and then use our
database to estimate the fraction of AGN as a function of the MIR flux and the \emph{F}(24 $\mu$m)/\emph{F}(\emph{R}) ratio.

Throughout this work with ``fraction of AGN'', we mean the fraction of extragalactic sources
in which is possible to reveal, via optical line width analysis and diagnostic diagrams (see, e.g., Veilleux \& Osterbrock 1987;
Kewley et al. 2006), the presence of an active nucleus, regardless of its strength relative to the host galaxy.
In the case of uncertain AGN signatures, we adopted a conservative approach.
Our database allows the first direct estimate (using optical spectra) of the fraction of AGN among the
sources with  large ($>$316) \emph{F}(24 $\mu$m)/\emph{F}(\emph{R}) ratios.

We adopt a flat cosmology with \emph{H$_0$} = 70 km s$^{-1}$ Mpc$^{-1}$,  $\Omega_M$=0.30, and
$\Omega_\Lambda$=0.70. Magnitudes are given in the Vega system. Unless otherwise stated, uncertainties are quoted at the 68\% (1$\sigma$) confidence
level.

%%%%%%%%%%%%%%%%%%%%%%%%%%%%%%%%%%%%%%%%%%%%%%%
\section{Multi-wavelength database}

The ELAIS-S1 field (center: $\alpha$=00$^h$35$^m$00$^s$.0, $\delta$=-43$^o$30$'$00$''$; J2000) is part of the SWIRE survey (Lonsdale et al. 2004) which is the largest {\it Spitzer} Legacy Program. SWIRE includes six high-latitude fields, totaling 49 deg$^2$ ($\sim$7 of which on the ELAIS-S1 region) observed by {\it Spitzer} in four bands (3.6, 4.5, 5.8, and 8.0 $\mu$m)
with the Infrared Array Camera (IRAC; Fazio et al. 2004), and in three bands (24, 70, and 160 $\mu$m) with the Multiband Imaging Photometer (MIPS; Rieke et al. 2004).

Originally the ELAIS-S1 field was selected as one of the four fields of the European Large Area {\it ISO} Survey (ELAIS) covering a total of 12 deg$^2$ at 15 $\mu$m (Oliver et al. 2000; Rowan-Robinson et al.  2004).  The size of the ELAIS-S1 field is 4 deg$^2$ and the 15 $\mu$m
catalog published by Lari et al. (2001) contains 329 extragalactic sources over
the flux range $0.5-100$ mJy.  The imaging and spectroscopic identification and classification of the
15$\mu$m sources have been presented by  La Franca et al. (2004; see also La Franca et al.  2007 and Gruppioni et al. 2008).

An area of 3.9 deg$^2$ is covered by radio observations obtained
with Australian Telescope Compact Array (ATCA) down to $S_{\rm
1.4-GHz} \simeq 30$ $\mu$Jy (1$\sigma$; Middelberg et al. 2008;
see also Gruppioni et al. 1999). ELAIS-S1 is also one of the
targets selected by the \emph{GALEX} Deep Survey, which has deeply
covered the central part of S1 in the far- and near-UV (Burgarella
et al. 2005).

%Fig 01
  \begin{figure}[]
  \centering
  \includegraphics[width=8.5cm]{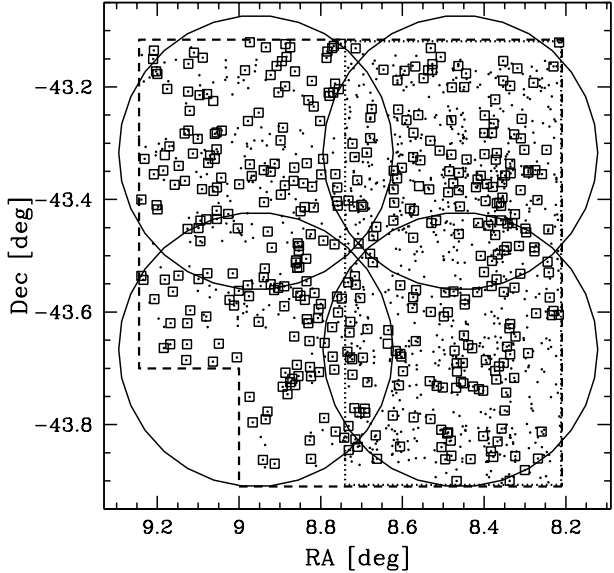}
  \caption{Position on sky of 3.6 $\mu$m (dots) and 24 $\mu$m
    (open squares) detected sources with a reliable redshift estimate.
The circles indicate the area covered by the four \emph{XMM-Newton}
pointings. The dashed and dotted lines show the statistically
useful areas of the 24 $\mu$m and 3.6 $\mu$m spectroscopic samples,
respectively (see Section 3.1).
             }
        \label{Fig_radecAllz}
  \end{figure}

%%%%%%%%%%%%%%%%%%%%%%%%%%%%%%%%%%%%%%%%%%%%%%%
\subsection{\emph{Spitzer} data}

The {\em Spitzer}/SWIRE observations in ELAIS-S1 cover a $\sim$7
deg$^2$ region larger than the whole 4 deg$^2$ {\em ISO} region,
in all the IRAC and MIPS bands, reaching 5$\sigma$
sensitivities of 3.7, 5.4, 48, 37.8, 230, 18$\times10^3$, and
150$\times10^3$ $\mu$Jy in the 3.6, 4.5, 5.8, 8.0, 24, 70, and 160
$\mu$m channels, respectively (Lonsdale et al. 2004). The SWIRE
data in ELAIS-S1 have been released to the community in autumn
2005 (Data Release 3, DR3) through the query building GATOR at the
NASA/Infrared Science Archive\footnote{\tt
http://irsa.ipac.caltech.edu/applications/Gator/}.
However, for the purposes of this work, we had access to the
SWIRE working catalogs, which contain all of the sources in the public
catalog, but reach deeper flux densities (on average about a factor of
30\% fainter). Details about
the SWIRE data reduction, that was carried out by the Spitzer
Science Center and SWIRE team, can be found in the Data Release
paper (Surace et al. 2005). Details on the generation of the {\em
Spitzer} band merged catalog can be found in Gruppioni et al.
(2008).

While the original SWIRE/ELAIS-S1 field is $\sim$7 deg$^2$, here we concentrate on a
central rectangular area, of 0.6 deg$^2$, with limits 8.21$^\circ$$<$$\alpha$$<$9.245$^\circ$ and
-43.91$^\circ$$<$$\delta$$<$-43.116$^\circ$ nearly including the four \emph{XMM-Newton} observations of the area described by Puccetti et al. (2006) (see Figure \ref{Fig_radecAllz}).
This area contains a total of 35021 sources detected by {\it Spitzer}, 32815 at 3.6 $\mu$m and
2056 at 24 $\mu$m  (1920 have been detected in both bands). Because of their bright flux limits, we excluded from the catalog the
MIPS measures at 70 and 160 $\mu$m.

%%%%%%%%%%%%%%%%%%%%%%%%%%%%%%%%%%%%%%%%%%%%%%%
\subsection{Optical and NIR  data}

$B$, $V$, and $R$  images,  down to $B\sim25$, $V\sim25$, and $R\sim24.5$ (95\% completeness) have been obtained with the
WFI at the ESO 2.2m telescope within the framework of the ESO-\emph{Spitzer} wide-area Imaging Survey
(ESIS; Berta et al. 2006; PI: Alberto Franceschini),
while about 1 deg$^2$ of ELAIS-S1 has been covered by deep $K^{'}$ and \emph{J} bands exposures with
SOFI at the ESO NTT telescope (I. Matute et al., in preparation).
Recently,  $I$- and \emph{z}-band photometry, down to $I\sim23$ and $z\sim22.5$ (90\% completeness), carried out with VIMOS at the Very Large Telescope (VLT),  has been released (Berta et al. 2008).

%%%%%%%%%%%%%%%%%%%%%%%%%%%%%%%%%%%%%%%%%%%%%%%
\subsection{X-ray band data}

The first very shallow ($\sim$10$^{-13}$ cgs; 2-10 keV) X-ray band observations on the ELAIS-S1 area were carried out with {\it BeppoSAX}
and presented by Alexander et al. (2001). More recently,
the central $\sim$0.6 deg$^2$ region of ELAIS-S1 has been surveyed in the X-ray band
with \emph{XMM-Newton} (four pointings of about 70 ``useful'' ks each) with 478 sources detected,
395 in the soft ($0.5-2$ keV) band down to a flux of $5.5 \times
10^{-16}$ cgs and 205 in the hard ($2-10$ keV) band down to a flux of $2 \times
10^{-15}$ cgs (Puccetti et al. 2006).

The regions with the highest \emph{XMM-Newton} sensitivity ($\sim$65\% of the full \emph{XMM-Newton} area) were later
target of 165 ks \emph{Chandra} observations, reaching on-axis sensitivities of $2 \times
10^{-15}$ cgs (2-10 keV; S. Puccetti et al., in preparation), with the aim of obtaining
precise positions for the X-ray sources.

%%%%%%%%%%%%%%%%%%%%%%%%%%%%%%%%%%%%%%%%%%%%%%%
\section{Optical spectroscopy}

ELAIS-S1 was the target of several spectroscopic campaigns with ESO telescopes.
The follow-up programs  of the 15 $\mu$m ELAIS-S1 sources
provided 60 identifications of R$<$22 sources within our investigated area (La Franca et al. 2004, 2007).
In the period 2004-2006, five follow-up programs were accepted, with the aim of obtaining spectroscopic identifications
for the \emph{XMM-Newton}, 3.6 and 24 $\mu$m sources in the SWIRE/\emph{XMM-Newton}/ELAIS-S1 area.
The number of spectra obtained in each spectroscopic run is shown
in Table \ref{table_runs}.

\begin{deluxetable}{c c c c c c c c}
\tabletypesize{\scriptsize}
%\rotate
\tablecaption{Number of good-quality\tablenotemark{a} spectra\label{table_runs}}
\tablewidth{0pt}
\tablehead{
\colhead{Period} & \colhead{Instrument} & \colhead{Total} & \colhead{3.6$\mu$m}& \colhead{24$\mu$m}  & \colhead{X\&24$\mu$m}  \\
}
\startdata
\nodata       & 15$\mu$m Catal.\tablenotemark{b} & \phantom{0}60 & \phantom{0}54 & \phantom{0}59 & \phantom{0}8 \\
73-04 & VIMOS     & 888 & 885 & 139 & 35\\
75-05 & VIMOS    & 328 & 327 & 129 & 18\\
76-05 & EFOSC2  & \phantom{0}20 & \phantom{0}19 & \phantom{0}12 &  \phantom{0}3\\
77-06 & FORS2-SLIT   & \phantom{0}11 & \phantom{0}11 & \phantom{0}11 & \phantom{0}6 \\
78-06 & FORS2-MOS    & \phantom{0}69 & \phantom{0}66 &\phantom{0}69 & 13 \\
\\
Total & \nodata & 1376 &1362 & 419 & 83 \\
\enddata
\tablenotetext{a}{\ We define as ``good quality'' those spectra with quality flag $\geq$1.5 (see section \ref{par:class}) }
\tablenotetext{b}{\ La Franca et al. (2004, 2007)}
\end{deluxetable}

%%%%%%%%%%%%%%%%%%%%%%%%%%%%%%%%%%%%%%%%%%%%%%%
\subsection{VIMOS observations}

Spectroscopic targets with a limiting magnitude of $R\sim24$ were observed with the VIsible MultiObject Spectrograph
(VIMOS) at the VLT in multi-object spectroscopy (MOS) mode, with the Low
Resolution Red (LRR) grism ($\lambda / \Delta \lambda \sim  210$), covering the
5500-9500 \AA\ wavelength range, with 1--4 hr exposure time per pointing, totaling
66 hr of observing time.
VIMOS data reduction was carried out using the VIMOS Interactive Pipeline
and Graphical Interface (VIPGI) developed by INAF Milano (Scodeggio et al. 2005).
Two observing runs were carried out in the years 2004 and  2005.
The first observing run was dedicated to identifying K-band and X-ray sources located in the western half
of the area, as the SWIRE
catalog was not yet available. The second run (as all the following runs)
was dedicated to observing 24 $\mu$m and X-ray sources,
while the 3.6 $\mu$m sources were used as fillers
of the masks. The observations were carried out
in the eastern half of the area, but unfortunately it was not possible to observe the southeastern
corner (see Figure \ref{Fig_radecAllz}). In total
1473 spectra were collected: 1212 of good quality (quality flag $\geq$1.5; see section \ref{par:class}
for a description of the quality flag).

%%%%%%%%%%%%%%%%%%%%%%%%%%%%%%%%%%%%%%%%%%%%%%%
\subsection{EFOSC2 and FORS2 (long slit) observations}

Four nights of visiting-mode observation in single-slit mode with EFOSC2 at the ESO-3.6m in
La Silla allowed us to observe 20 optically bright (R$<$20) targets
in the 4000-9000 \AA \ wavelength range with grisms 6 and 13.

Twenty hours of observing time was allocated with FORS2 at the VLT in long-slit mode. A total of  12 good-quality spectra
were obtained with the 150$I$ grism ($\lambda / \Delta\lambda\sim260$),
covering the wavelength range 3700-10300 \AA.

%%%%%%%%%%%%%%%%%%%%%%%%%%%%%%%%%%%%%%%%%%%%%%%
\subsection{FORS2 (MOS) observations}

Spectroscopy of 69 faint sources  has been carried out with FORS2 at the VLT in 2006
in MOS (MOvable-Slit) mode,
with 2.5 hr exposure times per pointing, corresponding to a total of  28 hr of observing time.
The same grism setup of the FORS2 Long-Slit observations was used (see the previous section).

%%%%%%%%%%%%%%%%%%%%%%%%%%%%%%%%%%%%%%%%%%%%%%%
\subsection{Data reduction}

The reduction process used standard MIDAS and IRAF facilities, except for
VIMOS data which required a specific software package (VIPGI; Scodeggio et al. 2005).
The raw data were bias-subtracted, corrected for pixel-to-pixel variations
(flat field), and eventually sky-subtracted.
Wavelength calibrations were carried out by comparison
with exposures of He and Ar lamps. Relative flux calibration was
carried out by observations of several spectrophotometric standard stars.

%%%%%%%%%%%%%%%%%%%%%%%%%%%%%%%%%%%%%%%%%%%%%%%
\subsection{Classification}\label{par:class}

Spectroscopic redshifts were obtained using both the {\tt rvidlines} IRAF tool and
an instrument-optimized MIDAS procedure, which allowed to compute
the average of the redshifts corresponding to each measured line center.
We assigned a quality flag to each redshift: [2] for the
determinations based on at least two confirmed features (typically 3-4 or more absorption/emission lines
besides the continuum shape), [1.5] for two very plausible features (besides the continuum shape),
[1] for one clearly recognizable feature, [0.5] for tentative estimates of redshift.
The spectroscopic catalogs we present include only highly reliable redshift estimates, with quality flag $\ge[1.5]$.

We classified the sources according to their optical
spectral features in five broad classes.

\begin{figure}[]
  \centering
  \includegraphics[width=8.5cm]{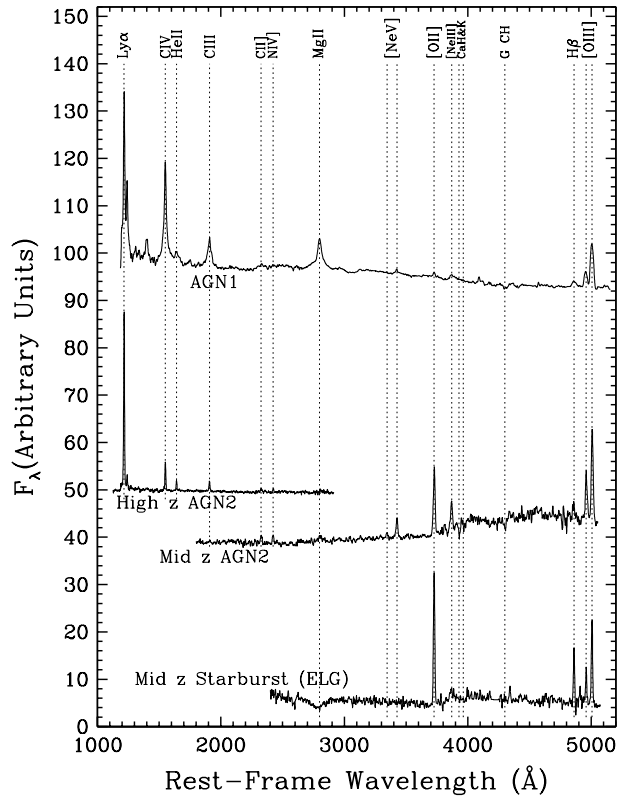}
     \caption{Composite spectra of 25 AGN1 (1$<$\emph{z}$<$3), 7 high (1.6$<$\emph{z}$<$2.6) and 12 middle (0.8$<$\emph{z}$<$1.2) redshift AGN2, and 20 ELG (0.6$<$\emph{z}$<$1.2).
             }
        \label{Fig_CompoSp}
  \end{figure}

   \begin{itemize}

\item Broad-line AGN (AGN1): type-1 AGN with broad
(FWHM $>$ 2000 km s$^{-1}$) emission lines such as \Civ, \Ciii,
\Mgii, \Hb, \Ha. As a reference, in  Figure
\ref{Fig_CompoSp},  the composite spectrum of a sub-sample of 25
(1$<$\emph{z}$<$3)\footnote{These redshift and wavelength intervals were
chosen in order to maximize the number of overlapping spectra.}
type-1 AGN is shown.

\item Narrow-line AGN (AGN2): type-2 AGN with narrow
(FWHM $<$ 2000 km s$^{-1}$) high-ionization emission lines
(\Civ, \Ciii, \Nev) or low-ionization (\Oii, \Hb, \OiiiD, \Ha, \Sii)
emission lines with flux ratios indicating the presence of an AGN (e.g., Osterbrock 1989;  Veilleux \& Osterbrock 1987; Tresse et al. 1996). In  Figure \ref{Fig_CompoSp},  the composite spectra of
seven high (1.6$<$\emph{z}$<$2.6) and 12 middle (0.8$<$\emph{z}$<$1.2) redshift$^{17}$ AGN2 are shown.

\item Emission-line galaxies (ELG): sources with narrow
emission lines, but no clear AGN signature in the optical spectra
(see Figure \ref{Fig_CompoSp}). They show strong low-ionization
emission lines that may be produced by thermal photons from hot
stars. They often show also \Cahk\ absorption and the continuum
break at 4000~\AA. This class is likely to include a fraction of
NL AGN,  possibly misclassified due to a small wavelength range
(e.g., only \Oii\ visible), low signal-to-noise ratio (S/N) spectra, or dilution by the host
galaxy emission. In  Figure \ref{Fig_CompoSp}, the composite
spectrum of 20 middle (0.6$<$\emph{z}$<$1.2) redshift$^{17}$ ELG is shown.

\item Absorption-line galaxies (GAL): sources with spectra
typical of early-type galaxies, characterized by absorption
features such as \Cahk\ and strong 4000~\AA\ continuum break.

\item Stars: sources with stellar spectra (mainly of cold M-stars),
characterized by a blackbody continuum and absorption features.

   \end{itemize}

%Fig 02

\begin{deluxetable*}{ccccccccccc}[h!]
\tabletypesize{\scriptsize}
\tablecaption{The spectroscopic catalog
\label{table_sp}} \tablewidth{0pt}
\tablehead{ \colhead{ID$_{\rm ESIS}$} & \colhead{R.A.} &
\colhead{Decl.} & \colhead{\it R} & \colhead{\emph{z}} &
\emph{q} & cls & ID$_{\rm SWIRE}$ & \emph{S}$_{\rm 3.6 \mu m}$ & \emph{S}$_{\rm 24 \mu m}$ & ID$_{\rm X}$ \\
~ &[deg] & [deg] & [mag]   & ~ &    ~    &       ~        & ~ &  [log($\mu$Jy)]      &  [log($\mu$Jy)]  & ~\\
(1) & (2)  & (3) & (4) & (5) & (6) & (7) & (8) & (9) & (10) & (11)}
\startdata
     J003522.49-432517.45 & 8.84371 & -43.42152 & 18.65 & 0.225 & 2.0 & 4 & SWIRE3\_J003522.50-432517.3 & 2.2335 & \nodata &  311 \\
     J003521.34-432509.51 & 8.83894 & -43.41928 & 21.71 & 0.636 & 2.0 & 2 & SWIRE3\_J003521.36-432509.3 & 1.7075 & \nodata &  301 \\
     J003453.39-431802.46 & 8.72240 & -43.30064 & 20.65 & 1.075 & 2.0 & 1 & SWIRE3\_J003453.39-431802.1 & 1.8465 & 2.5470  &  258 \\
     J003443.36-431713.39 & 8.68062 & -43.28707 & 18.42 & 0.000 & 2.0 & 5 & SWIRE3\_J003443.36-431713.3 & 2.7933 & \nodata &  245 \\
     J003407.39-430751.01 & 8.53081 & -43.13084 & 18.91 & 0.188 & 2.0 & 2 & SWIRE3\_J003407.41-430750.9 & 2.2723 & 2.1869  &  185 \\
     J003424.30-432037.21 & 8.60127 & -43.34372 & 21.47 & 1.041 & 2.0 & 1 & SWIRE3\_J003424.32-432037.2 & 1.5117 & 2.5663  &  216 \\
     J003429.29-432409.80 & 8.62204 & -43.40269 & 21.28 & 1.065 & 2.0 & 1 & SWIRE3\_J003429.30-432409.5 & 1.7462 & \nodata &  229 \\
     J003416.03-433338.06 & 8.56669 & -43.56065 & 22.17 & 0.956 & 2.0 & 1 & SWIRE3\_J003416.02-433338.2 & 1.6669 & 2.5285  &  205 \\
     J003320.65-433716.86 & 8.33610 & -43.62132 & 19.48 & 0.287 & 2.0 & 2 & SWIRE3\_J003320.68-433716.6 & 2.1074 & 2.4607  &   80 \\
     J003503.93-432847.13 & 8.76638 & -43.47975 & 20.78 & 1.108 & 2.0 & 1 & SWIRE3\_J003503.94-432846.9 & 2.2380 & 2.8999  &  270 \\
     J003716.73-434151.12 & 9.31958 & -43.69757 & 17.91 & 0.226 & 2.0 & 3 & SWIRE3\_J003716.71-434151.1 & 2.5961 & 3.3699  & \nodata \\
     J003546.69-430340.11 & 8.94461 & -43.06111 & 17.30 & 0.147 & 2.0 & 3 & SWIRE3\_J003546.72-430339.8 & 2.8383 & 3.4389  & \nodata \\
     J003635.09-430133.70 & 9.14621 & -43.02603 & 17.61 & 0.208 & 2.0 & 3 & SWIRE3\_J003635.10-430133.6 & 2.7766 & 3.4651  & \nodata \\
     J003531.00-430117.55 & 8.87917 & -43.02154 & 17.14 & 0.146 & 2.0 & 3 & SWIRE3\_J003531.01-430117.4 & 2.7604 & 3.3664  & \nodata \\
\enddata
\tablecomments{Column 1: ESIS identification name from Berta et al. (2006). Columns 2 and 3: coordinates (J2000). Column 4: \emph{R}-band Vega magnitude.  Column 5: redshift.
Column 6: redshift quality code: [2.0]=reliable, based on $>$2 confirmed lines; [1.5]=very plausible, based on two lines. Column 7:
spectroscopic classification code: [1]=type-1 AGN; [2]=type-2 AGN; [3]=ELG; [4]=normal galaxy; [5]=star. Column 8:  SWIRE identification name
of the most probable corresponding {\em Spitzer} source.
Column 9: flux at 3.6 $\mu$m.  Column 10: flux at 24 $\mu$m.  Column 11: \emph{XMM-Newton} source name, see Feruglio et al. (2008).\\
(This table is available in its entirety in a machine-readable form in the
online edition of the {\it Astrophysical Journal}.  A portion is
shown here for guidance regarding its form and content.)
}
\end{deluxetable*}

The wavelength range of our spectra (mostly in the interval
4500-9500~\AA), is wide enough to classify the sources. In the case of
those narrow lines sources,  where only the region of the \Oii\
line was visible, we used the presence (or not) of the
high-ionization \Nev\ line to discriminate between AGN2 and ELG.
In any case, the AGN2 classification has been conservatively
assigned only in fully reliable conditions.
 Because of the presence of broad emission lines in their spectra,
even in the case of low S/N data,
we can exclude that type-1 AGN have been wrongly classified
as either AGN2 or ELG.

As far as the galaxies with no emission lines (passive) are
concerned, we were not able to measure redshifts larger than one
(with calcium-break redshifted out of the wavelength coverage).
However, this bias should be small in the 24 $\mu$m sample, as the
fraction of energy emitted in the MIR by the passive galaxies is
smaller than in the other galaxies (see, e.g., the SED library
shown in Figure 1 of Polletta et al. 2007), and therefore the 24
$\mu$m sample should be scarcely populated by no emission line
galaxies (see discussion in Section 4.1). This bias could instead
affect the 3.6 $\mu$m sample\footnote{Where indeed we found a
fraction of classified no emission line galaxies four times larger
than in the 24 $\mu$m sample; see discussion in Section 4.2.} which,
however, has not been scientifically used in this work.

%%%%%%%%%%%%%%%%%%%%%%%%%%%%%%%%%%%%%%%%%%%%%%%
\section{The spectroscopic catalog}

In Table \ref{table_sp}, we present the whole spectroscopic catalog consisting of 1376 optical counterparts
of 3.6 and 24 $\mu$m {\it Spitzer} sources (whose
redshift estimates have a quality flag $\ge[1.5]$).
For every source we report: the ESIS catalog identification code (from Berta et al. 2006),
sky coordinates in degree units (J2000) of the optical counterpart and its
R-band magnitude, redshift,  redshift quality code (only sources with
1.5 or 2.0 redshift quality are included), spectroscopic classification code ([1]=type-1
AGN; [2]=type-2 AGN; [3]=ELG; [4]=normal galaxy; [5]=star). We report also the SWIRE identification code and the 3.6 and 24
$\mu$m flux densities, in log[$\mu$Jy] units,
of the corresponding {\em Spitzer} sources. For those sources detected in the X-rays, the ID code
from the catalog of Feruglio et al. (2008) is also reported.

The spectroscopic sample of the (both {\em Spitzer }detected and not)
\emph{XMM-Newton}  sources has been presented in a separate publication by Feruglio et al. (2008).
Therefore, the spectroscopic identification of any X-ray detected {\em Spitzer} source appears in both catalogs.

%%%%%%%%%%%%%%%%%%%%%%%%%%%%%%%%%%%%%%%%%%%%%%%
\subsection{The 24 $\mu$m sample}\label{par:24umsample}

Because of the incomplete execution of the second VIMOS run, the statistically useful area of the field for spectroscopic
identifications of the  24 $\mu$m sources is a rectangle with limits 8.21$^\circ$$<$$\alpha$$<$9.245$^\circ$ and -43.91$^\circ$$<$$\delta$$<$-43.116$^\circ$, where all sources having $\alpha$$>$9.0$^\circ$ and $\delta$$<$-43.7$^\circ$ have been excluded (see Figure \ref{Fig_radecAllz}).

%Fig 03
  \begin{figure}[]
  \centering
  \includegraphics[height=8.5cm, angle=-90]{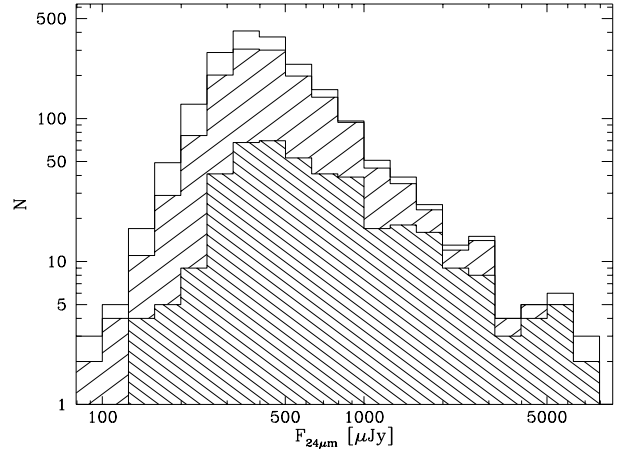}
     \caption{24 $\mu$m flux density distribution of all {\em Spitzer} sources  (line on top),
     those with an \emph{R}-band detection (hatched region) and
     those with a reliable redshift estimate (darker shaded area).
             }
        \label{Fig_histo_F24}
  \end{figure}

%Fig 04
  \begin{figure*}[t]
  \centering
  \includegraphics[width=11cm, angle=-90]{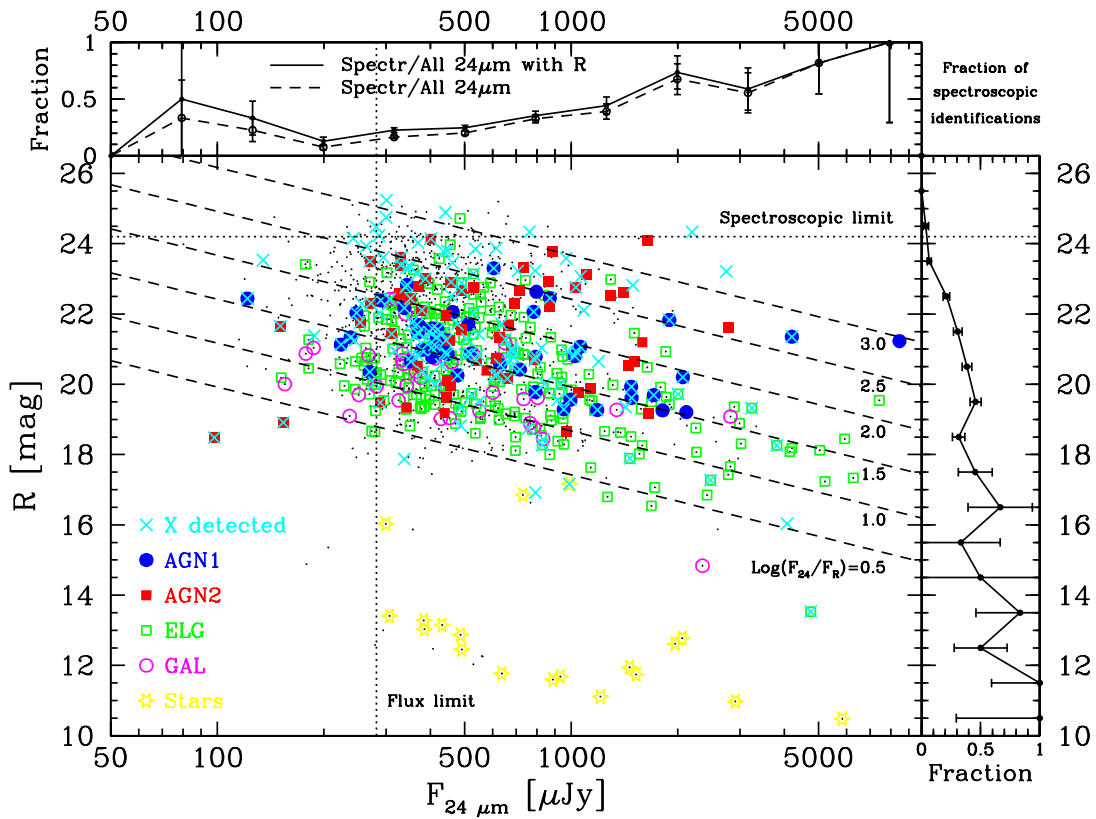}
     \caption{Central: \emph{R} magnitude as a function of the 24 $\mu$m flux.
      Those having a
     reliable redshift estimate are represented by symbols whose meaning is
     shown in the legend. Sources detected in the X-ray are represented by cyan crosses.
    The dashed lines indicate the loci with equal \emph{F}(24 $\mu$m)/\emph{F}(\emph{R}) ratio (whose values are reported in the labels).
    The vertical dotted line represents the 24 $\mu$m flux limit  of 280 $\mu$m used to measure the fraction of AGN,
    while the horizontal dotted line represents the \emph{R}=24.2 mag limit of the spectroscopic identifications.
      Upper: fraction of the spectroscopically identified sources as a function of the 24 $\mu$m flux
     for the whole 24 $\mu$m sample (dashed line) and the \emph{R}-band detected 24 $\mu$m sample (continuous line).
     Right: fraction of the spectroscopically identified 24 $\mu$m sources as a function of the \emph{R}-band magnitude.
             }
        \label{Fig_RvsF24}
  \end{figure*}

A total of 1932 24 $\mu$m sources were detected by {\it Spitzer}
within this 0.56 deg$^2$ studied area, 1512 of which (78\%)
have been identified in the R-band optical  ESIS catalog. 419
sources have been spectroscopically identified: 399 result to be
extragalactic sources and 20 stars. The extragalactic sources were
classified into 107 AGN (27.3\%, 52 AGN1 and 57 AGN2), 253 ELG
(63.4\%), and 37 GAL (9.3\%). Figure \ref{Fig_histo_F24} shows the
histogram of the 24 $\mu$m flux of all sources (top line),
those with an R-band detection (hatched region) and those with a
reliable redshift estimate (darker shaded area).

In Figure \ref{Fig_RvsF24}, the distribution of the $R$ magnitude as a function of the 24 $\mu$m flux
is shown, with the indication of the spectroscopically identified and the X-ray detected
sources.  The fraction of spectroscopically identified sources decreases at both
faint optical and 24 $\mu$m fluxes. This trends should be taken into account when these data are used.
In our analyses (see section \ref{cap:fractionAGN}), we selected a sub-sample with 24 $\mu$m fluxes
brighter than 280 $\mu$Jy and magnitudes brighter than R=24.2,  which corresponds to the largest
possible sample where sufficient spectroscopic information exists (at these flux limits the fraction of spectroscopically identified sources is about 10\%).

\begin{figure}[]
  \centering
  \includegraphics[height=9.0cm, angle=-90]{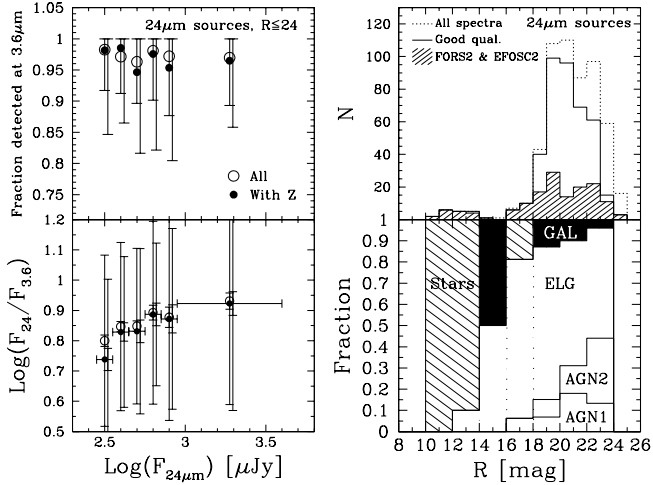}
     \caption{Upper left: fraction of the 24 $\mu$m sources brighter than \emph{R}=24
     which have been detected at 3.6 $\mu$m as a function of the 24 $\mu$m flux.
     The whole sample and the spectroscopically
     identified one are represented by open and filled circles, respectively.
    Lower left: ratio of the 24 $\mu$m and 3.6 $\mu$m flux densities
    as a function of the  24 $\mu$m flux for the same samples. Symbols are
    as in the previous panel.  Thick vertical error bars correspond to the (1$\sigma$) uncertainties on the mean, while
    the light vertical error bars show the (1$\sigma$) spread of the distributions.
    Upper right: \emph{R} magnitude distribution of the spectroscopically
     observed 24$\mu$m sources. The dotted line is the distribution of all the spectra, while
     the continuous line is the distribution of the ``good-quality'' (qual$\geq$1.5) spectra presented in this work.
     The shaded area is the histogram of those sources whose spectra have been mainly obtained with FORS2/VLT (some of the brightest with  EFOSC2/3.6\ m or DFOSC). Lower right: fraction of the  ``good-quality'' (qual$\geq$1.5) spectra of the 24$\mu$m sources
     according to the spectroscopic class. Each histogram includes the classes below itself.
             }
        \label{Fig_Col24}
  \end{figure}

We have compared the spectroscopic sample (including only sources with a reliable redshift identification; see the
previous section) with the ``parent'' 24 $\mu$m sample.
In the upper left panel of Figure \ref{Fig_Col24}, the fraction of the 24 $\mu$m sources brighter than $R$=24 mag which have been detected at 3.6 $\mu$m as a function of the 24 $\mu$m flux is shown.
At 24 $\mu$m fluxes brighter than 280 $\mu$Jy (once the uncertainties are taken into account),
no significant difference
is found between the spectroscopic sample and the parent 24 $\mu$m sample: in both samples, in any bin
more than 95\% of the sources brighter than $R$=24
have been detected at 3.6 $\mu$m. No significant difference is found, either, between the two samples, in the
value of the \emph{F}(24 $\mu$m)/F(3.6 $\mu$m) ratio as a function of the 24 $\mu$m flux (see the lower left panel in Figure \ref{Fig_Col24}).
We can then affirm that, although the fraction of spectroscopically
identified sources depend either from the optical and 24 $\mu$m fluxes,
the average MIR SED properties of the spectroscopically identified sources
are not significantly different from those of the parent
24 $\mu$m sample (with $R$ brighter than 24 mag).

%Fig 05

 In the upper right panel of Figure \ref{Fig_Col24}, the \emph{R}-band distribution of the spectroscopically
 identified 24 $\mu$m sources is shown. Unlike the identifications carried out with FORS2 and EFOSC2,
where almost all spectra down to  $R\sim24$ mag are useful (quality-flag $\geq$1.5), at \emph{R}-band magnitudes fainter than $R\sim19$ mag
 the fraction of low-quality VIMOS spectra increases with decreasing optical fluxes: it is about 20\%, 35\%, and 85\% in the
21-22, 22-23, and 23-24 \emph{R}-band magnitude bins, respectively. As a consequence, most ($\sim$80\%) of the spectroscopic
identifications at magnitudes fainter than \emph{R}=23 mag have been carried out only with FORS2, which is indeed much more efficient
(because of better sky and fringe subtraction) at faint magnitudes.
In the lower right panel of Figure \ref{Fig_Col24}, the \emph{R}-band magnitude distribution of the fraction of each spectroscopic
class among the good-quality spectra is shown.

%Fig 06
  \begin{figure}[]
  \centering
  \includegraphics[width=8.5cm, angle=0]{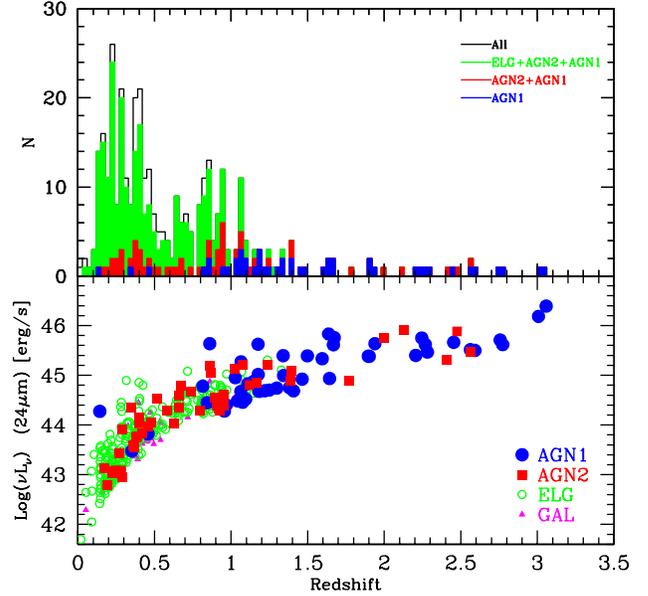}
     \caption{Top: redshift distribution of all 24 $\mu$m
     detected {\em Spitzer} sources. Those spectroscopically classified as AGN1, AGN2, and ELG
     are indicated in blue, red, and green colors, respectively. Bottom:
     log($\nu$L$_\nu$) 24 $\mu$m luminosity--redshift diagram  of all spectroscopically identified 24 $\mu$m
     sources. Blue circles, red squares, and green circles  represent AGN1, AGN2, and ELG, respectively. }
        \label{Fig_L24_z}
  \end{figure}

The redshift histogram of the spectroscopically identified 24
$\mu$m sources and  the 24 $\mu m$ luminosity--redshift diagram are
shown in Figure \ref{Fig_L24_z}.  The luminosities have been
computed applying a \emph{k}-correction derived by a best SED fitting (F. La
Franca et al., in preparation):  following Fiore et al. (2008) we used a
modified version of the  SED library  described in Polletta et al.
(2007) and already used for similar analyses by Lonsdale et al.
(2004), Franceschini et al. (2005), Hatziminaoglou et al. (2005),
Polletta et al. (2006), Weedman et al.
(2006), and Tajer et al. (2007). The luminosity--redshift
distribution is that typical of a flux-limited sample. On average,
the most distant and luminous sources are the AGN1, then there are
the AGN2, and finally the ELG. As this is also an optical limited
sample (the spectroscopic identifications reach $R\sim24.2$ mag),
the explanation of this behavior requires a proper analysis (which
is beyond the scope of this paper) of the bivariate MIR/optical
luminosity functions of each population, combined with the
\emph{k}-corrections and all the selection effects. However, several
works have been  dedicated to the understanding of the shape of
the 24 $\mu$m redshift distribution, discussing the combination of
the effects of both the luminosity functions evolution and the
MIR/optical SEDs (see, e.g., P\'erez-Gonz\'alez et al. 2005, Caputi et
al. 2006, Desai et al. 2008). Desai et al. (2008) discuss in detail
the redshift distribution of a sample of 591 24 $\mu$m sources,
detected in the Bo\"otes field of the NOAO Deep Wide-Field Survey,
down to   \emph{F}(24 $\mu$m) = 300 $\mu$Jy and spectroscopically
identified down to \emph{R}=25 mag (thus with flux limits similar to our
sample). They obtained optical spectroscopic redshift for 71\% of
the sources. Similarly to our sample, their redshift distribution
shows a peak at $z\sim0.3$ and a possible additional peak at
$z\sim0.9$ (see their Figure 9).  These peaks could be attributed
to redshifted emission features in the SED which enter the 24
$\mu$m MIPS bandpass. According to the analysis of Desai et al.
(2008), the  $z\sim0.3$ peak is difficult to be reproduced, as
only the polycyclic aromatic hydrocarbon (PAH) emission features at
16.3 and 17 $\mu$m can be partly responsible of its presence,
while the $z\sim0.9$ peak could be attributed to the 12.7 $\mu$m
PAH feature and the 12.8 $\mu$m [Ne~{\tiny II}] emission line
passing trough the 24 $\mu$m bandpass. However, it is interesting
to note that these two peaks are fairly well reproduced by the
galaxy evolution model of Lagache et al. (2004) (see Figure 11 in
Desai et al. 2008) where both normal and starburst galaxies are
represented by luminosity-dependent SEDs. The model encompasses a
strong rate of evolution of the luminosity density of starburst
galaxies, peaking at $z\sim0.7$ and remaining constant up to
\emph{z}=4, while the normal galaxies evolve up to \emph{z}=0.4, after which
their luminosity density remains constant. Analyzing a fraction of
their sources without emission-line redshifts, Desai et al. (2008)
find also weak evidence for another peak at $z\sim2$ where about
55\% of the sources are AGN-dominated.

At redshift larger than 1.0, our distribution is populated by AGN.
AGN1 show larger redshifts and luminosities than AGN2:  80\% of
the AGN1 have redshift larger than $\sim$0.9, while 80\% of the
AGN2 show redshift lower than $\sim$1.1. This redshift
distribution is not unusual for an MIR sample combined to an
optical flux limit (introduced by the spectroscopic identification
and classification process). See, e.g.,  the luminosity--redshift
distribution of the 15 $\mu$m selected (F(15 $\mu$m)$>$500
$\mu$Jy) sample in the whole ELAIS-S1 region by Matute et al.
(2006; their Figure 3) where the spectroscopic identifications have
been carried out down to $R\sim22$ mag. In the 15 $\mu$m sample
used by Matute et al. (2006), 90\% of the AGN1 show redshift larger
than $\sim$0.4, while 80\% of the AGN2 have redshift lower than
$\sim$0.4. The difference between the redshift distribution of the
two AGN populations is due to a combination of  their luminosity
functions with a difference in the average MIR/Optical ratio of
their SED. Although both AGN1 and (with larger uncertainties) AGN2
show a strong luminosity evolution in the MIR (e.g., Matute et al.
2006 find a luminosity evolution at 15 $\mu$m of  $L(z)=L(0)\times
(1+z)^k$ with $k\sim2.9$ for AGN1 and $k\sim1.8-2.6$ for AGN2; see
also Brown et al. (2006) for a recent estimate of the  AGN1
evolution at 24 $\mu$m),  AGN2 have on average larger \emph{F}(24 $\mu$m)/\emph{F}(\emph{R}) ratios than AGN1 (see, e.g., La Franca et al. (2004) and
the discussion in Section 5.1.1 of this paper). In our sample,
about 40\% of the AGN2 show F(24$\mu$m)/\emph{F}(\emph{R}) ratios larger than
100, while only 15\% of the AGN1 have F(24$\mu$m)/\emph{F}(\emph{R}) ratios
larger than this limit. A similar kind of difference between AGN1
and AGN2 is seen in the X-rays, as far as the X-ray to optical
ratio is concerned (see, e.g., Fiore et. al. 2003). This happens
because, in  AGN2,  the optical AGN component is more often
obscured, and the resulting optical spectrum is the combination of
the hosting galaxy emission with the AGN Narrow Line Region (when
visible; see, e.g., Fiore et al. (2000, 2003), Cocchia et al. (2007), and
Caccianiga et al. (2008) for a discussion on the X-ray Bright
Optically Normal Galaxies--XBONG). The average lower optical
luminosity of AGN2 compared to the AGN1 (with the same 24 $\mu$m
luminosity), combined to the optical R=24.2 mag spectroscopic
limit, is therefore one of the most important agent for the
average lower redshifts of the AGN2 sample (the luminosity
function and the \emph{k}-corrections are the other most important
ingredients). In summary, as the measure of the 24 $\mu$m AGN1 and
AGN2  luminosity function is beyond the scope of this paper, we
can comment that the similarity between the AGN1 and AGN2
luminosity--redshift distribution of our sample and that observed
at 15 $\mu$m (Matute et al. 2006), suggests that it is
qualitatively compatible with the previous estimates of the AGN
evolution in the MIR (e.g., Matute et al. 2002, 2006; see also
Brown et al. 2006), where all these selection effects have been
taken into account.

%%%%%%%%%%%%%%%%%%%%%%%%%%%%%%%%%%%%%%%%%%%%%%%
\subsection{The 3.6 $\mu$m sample}

Although the spectroscopic follow up was dedicated to the identification of the 24 $\mu$m and X-ray sources,
a total of 1362 3.6 $\mu$m sources (881 not detected either in the 24 $\mu$m or the X-ray bands) have been
spectroscopically identified. These spectra come from the identifications of 3.6 $\mu$m sources used as
fillers of the VIMOS and FORS2 masks, or from the identification of the \emph{K}-band (2.2 $\mu$m) sources of the
western half area, observed during the first VIMOS run (see Section 3).
We limit the following discussion to the western half of the VIMOS area
having limits 8.21$^\circ$$<$$\alpha$$<$8.74$^\circ$ and -43.91$^\circ$$<$$\delta$$<$-43.116$^\circ$
that can be treated as an unbiased representation of the parent 3.6 $\mu$m sample and
in which most of these sources (999) are indeed located (see Figure \ref{Fig_radecAllz}).

%Fig 07
  \begin{figure}[]
  \centering
  \includegraphics[height=8.5cm,angle=-90]{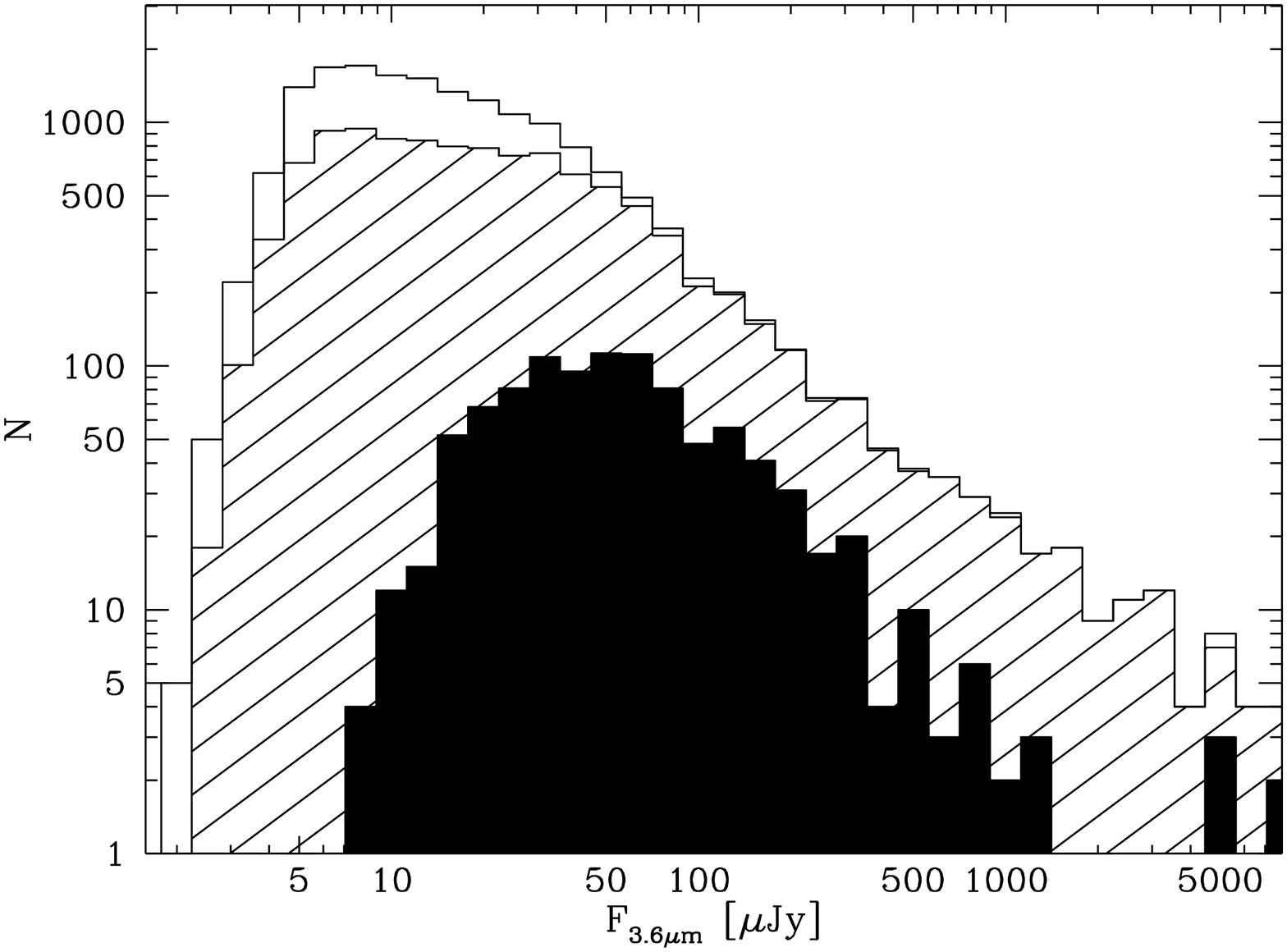}
     \caption{3.6 $\mu$m flux distribution of all {\it Spitzer} sources (line on top),
     those with an \emph{R}-band detection (hatched region), and
     those with a reliable redshift estimate (darker shaded area).
             }
        \label{Fig_histo_F36}
  \end{figure}

A total of 16782 3.6 $\mu$m sources were detected by {\it Spitzer} within this 0.31 deg$^2$ area, 10794 of which
(64\%) have been identified in the \emph{R}-band optical ESIS catalog. 999 sources have been spectroscopically identified:
762 result to be extragalactic sources and 237 stars. The
extragalactic sources were classified into 111 AGN (14.6\%, 51 AGN1
and 60 AGN2), 358 ELG (47.0\%), and 293 GAL (38.5\%). Figure
\ref{Fig_histo_F36} shows the 3.6 $\mu$m flux  histogram of the
sub-sample (line on top), those with an \emph{R}-band detection (hatched
region) and those with a reliable redshift estimate (darker shaded
area). The fraction of 3.6 $\mu$m sources with spectroscopic
redshift is 22\% for sources with \emph{F}(3.6 $\mu$m)$>$50 $\mu$Jy and
decreases at fainter fluxes ($\sim$1\%  at \emph{F}(3.6 $\mu$m)$\sim$30
$\mu$Jy).

%Fig 08
  \begin{figure*}[]
  \centering
  \includegraphics[width=11cm,angle=-90]{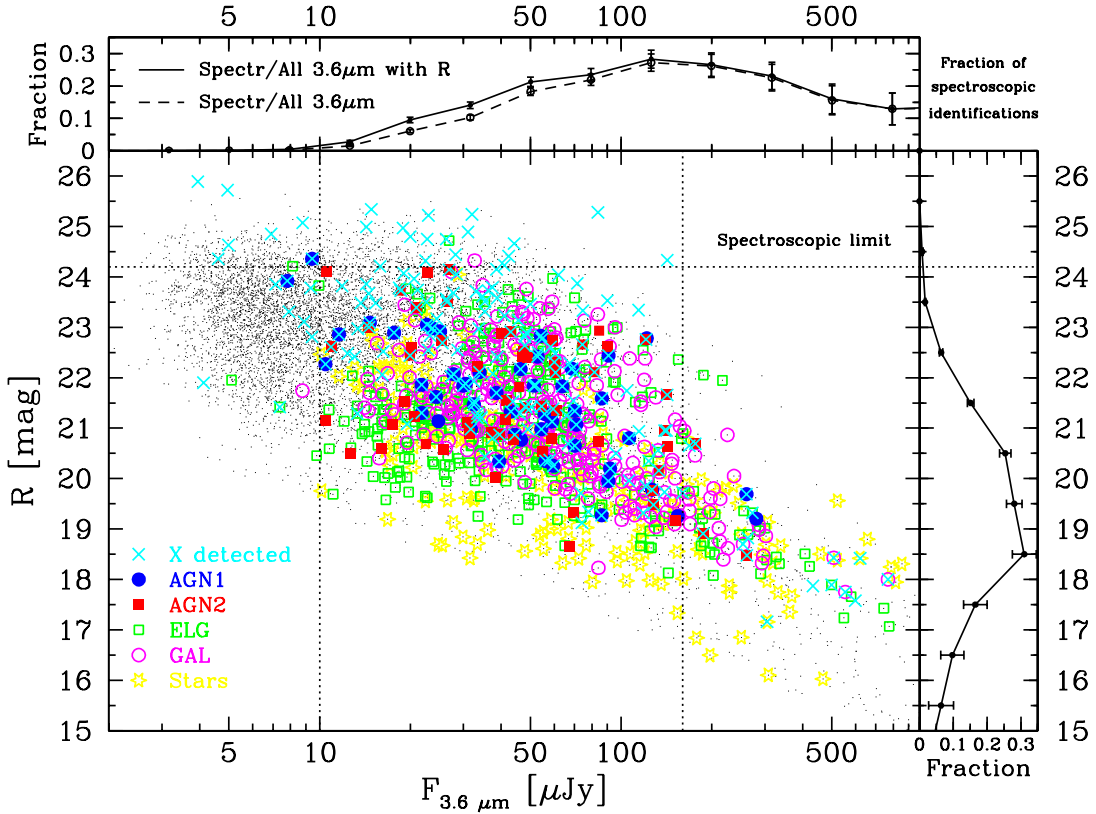}
      \caption{Central: \emph{R} magnitude as a function of the 3.6 $\mu$m flux of the sources in the western half of the area.
      Those having a
     reliable redshift estimate are represented by symbols whose meaning is
     shown in the legend. Sources detected in the X-ray are indicated with cyan crosses. The two dotted vertical lines
     show the flux limits where, in each \emph{R}-band magnitude interval, the spectroscopic identifications are a
    fair (``gray'') sample of the 3.6 $\mu$m catalog (see Section 4.2). The horizontal dotted line represents the \emph{R}=24.2 mag limit of the spectroscopic identifications.
     Upper: fraction of the spectroscopically identified sources as a function of the 3.6 $\mu$m flux
     for the whole 3.6 $\mu$m sample (dashed line) and the \emph{R}-band detected 3.6 $\mu$m sample (continuous line).
     Right: fraction of the spectroscopically identified 3.6 $\mu$m sources as a function of the \emph{R}-band magnitude.
             }
        \label{Fig_RvsF36}
  \end{figure*}

In Figure \ref{Fig_RvsF36}, the distribution of the \emph{R} magnitude as a function of the 3.6 $\mu$m flux
is shown, with the indication of the spectroscopically identified and the X-ray detected
sources.  The fraction of spectroscopically identified sources decreases at both
faint optical and 3.6 $\mu$m fluxes. Unlike the 24 $\mu$m sample, a decrease in the fraction
of identifications is evident at bright optical and 3.6 $\mu$m fluxes too (see also Figure \ref{Fig_histo_F36}).
This is due to the fact that the spectroscopic identifications of the sources brighter than \emph{R}=18 were carried out
by single slit observations of the 24 $\mu$m sources only.

%Fig 09
  \begin{figure}[]
  \centering
  \includegraphics[height=9.0cm,angle=-90]{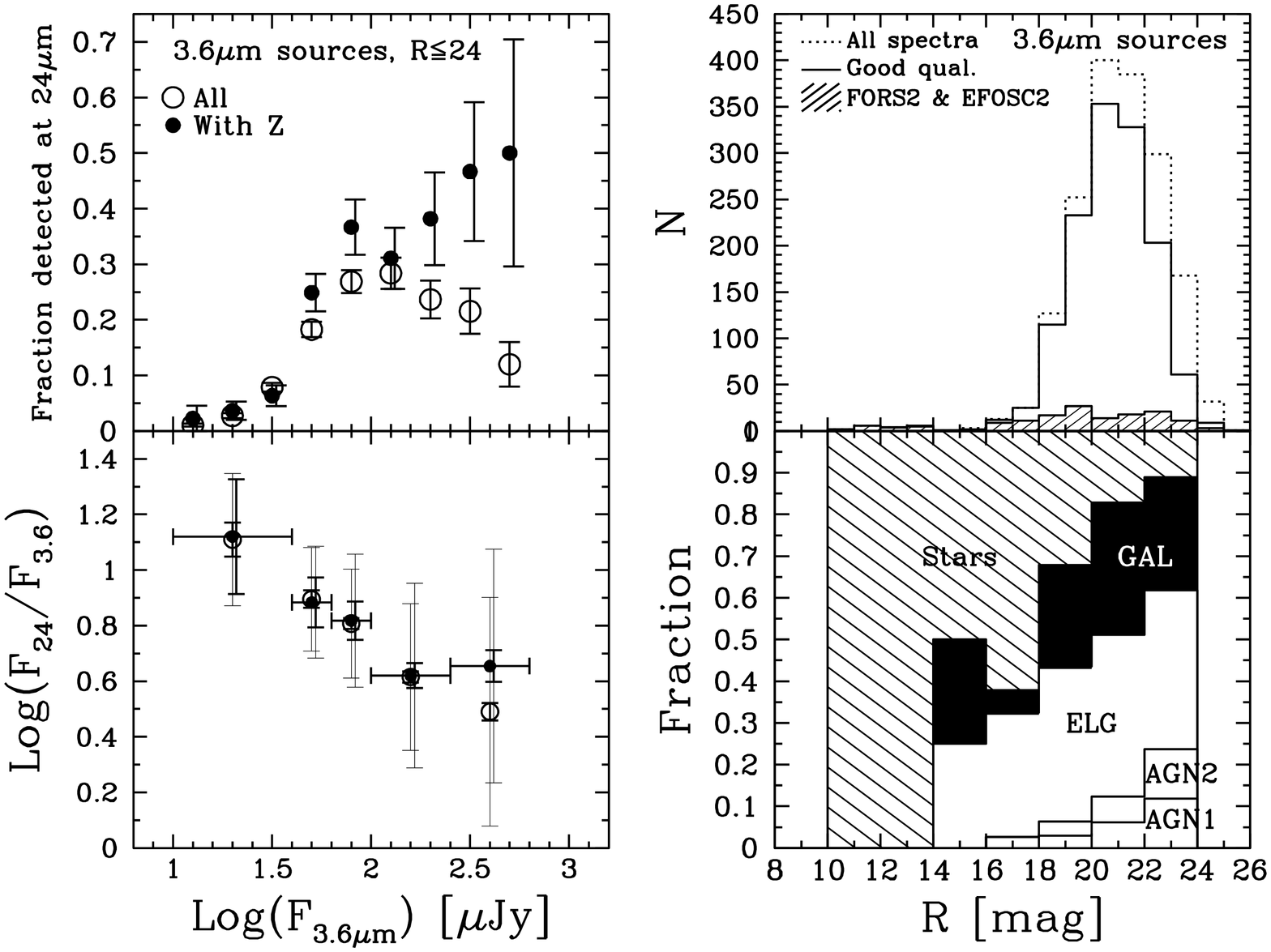}
     \caption{Upper left: fraction of the 3.6 $\mu$m sources brighter than \emph{R}=24
     which have been detected at 24 $\mu$m as a function of the 3.6 $\mu$m flux.
     The whole sample and the spectroscopically
     identified one are represented by open and filled circles, respectively.
    Lower left: ratio of the 24 $\mu$m and 3.6 $\mu$m flux densities
    as a function of the  3.6 $\mu$m flux for the same samples. Symbols are
    as in the previous panel.  Thick vertical error bars correspond to the (1$\sigma$) uncertainties on the mean, while
    the light vertical error bars show the (1$\sigma$) spread of the distributions. Upper right: \emph{R} magnitude distribution of the spectroscopically
     observed 3.6 $\mu$m sources. The dotted line is the distribution of all the spectra, while
     the continuous line is the distribution of the ``good-quality'' (qual$\geq$1.5) spectra presented in this work.
     The shaded area is the histogram of those sources whose spectra have been mainly obtained with FORS2/VLT (some of the brightest with  EFOSC2/3.6\ m or DFOSC). Lower right: fraction of the  ``good-quality'' (qual$\geq$1.5) spectra of the 3.6 $\mu$m sources
     according to the spectroscopic class. Each histogram includes the classes below itself.
             }
        \label{Fig_Colore36}
  \end{figure}

This bias is evident in the upper left panel of Figure \ref{Fig_Colore36}  where the fraction of the 3.6 $\mu$m sources brighter than \emph{R}=24
which have been detected at 24 $\mu$m as a function of the 3.6 $\mu$m flux is shown.
In the 3.6 $\mu$m flux interval between 10 and 160 $\mu$Jy, there is no difference between the spectroscopic sample and the
parent  3.6 $\mu$m sample. No difference is found either, in the same flux interval, between  the \emph{F}(24 $\mu$m)/\emph{F}(3.6 $\mu$m) ratios of
the two samples  (see the lower left panel in Figure \ref{Fig_Colore36} ).  On the other hand, at 3.6 $\mu$m fluxes brighter than 160 $\mu$Jy  the
fraction of 24 $\mu$m detected spectroscopically identified sources is significantly larger than the parent 3.6 $\mu$m sample.
We can then conclude that, within the western half of the area,  in each \emph{R}-band magnitude interval brighter than $R\sim24$, the 3.6 $\mu$m spectroscopic catalog is a fair  sample of the 3.6 $\mu$m sources only in the 10-160 $\mu$Jy flux range.

\begin{figure}[]
  \centering
  \includegraphics[height=8.5cm,angle=-90]{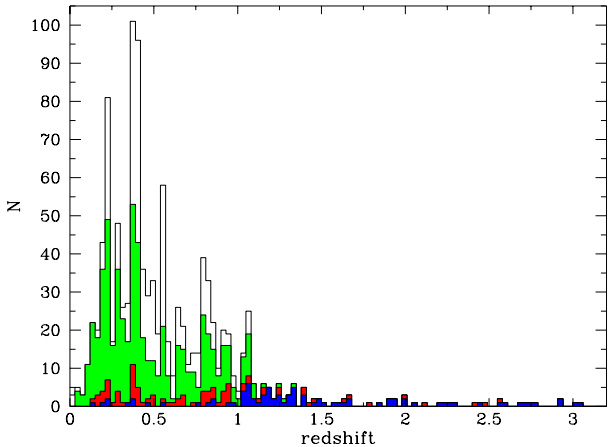}
     \caption{Redshift distribution of the 3.6 $\mu$m
     detected {\em Spitzer} sources. Those spectroscopically classified as AGN1, AGN2, and ELG
     are indicated in blue, red, and green colors, respectively (as in Figure 6).
             }
        \label{Fig_histo_z_36}
  \end{figure}

 In the upper right panel of Figure \ref{Fig_Colore36}  the \emph{R}-band distribution of the spectroscopically
 identified 3.6 $\mu$m sources is shown, while
in the lower right panel of Figure \ref{Fig_Colore36}  the \emph{R}-band magnitude distribution of the fraction
of each spectroscopic class among the good-quality spectra is shown. Unlike the
spectroscopic 24 $\mu$m sample (see Figure \ref{Fig_Col24} for comparison) even at faint \emph{R}-band magnitudes ($R\sim22-24$), a large fraction (about 40\%) of the sources are non-emission line galaxies (GAL) and stars.
The redshift histogram of the spectroscopically identified 3.6 $\mu$m sources is shown
in Figure \ref{Fig_histo_z_36}.

%Fig 10

%%%%%%%%%%%%%%%%%%%%%%%%%%%%%%%%%%%%%%%%%%%%%%%
\subsubsection{IRAC colors}

\begin{figure}[]
  \centering
  \includegraphics[width=8.5cm]{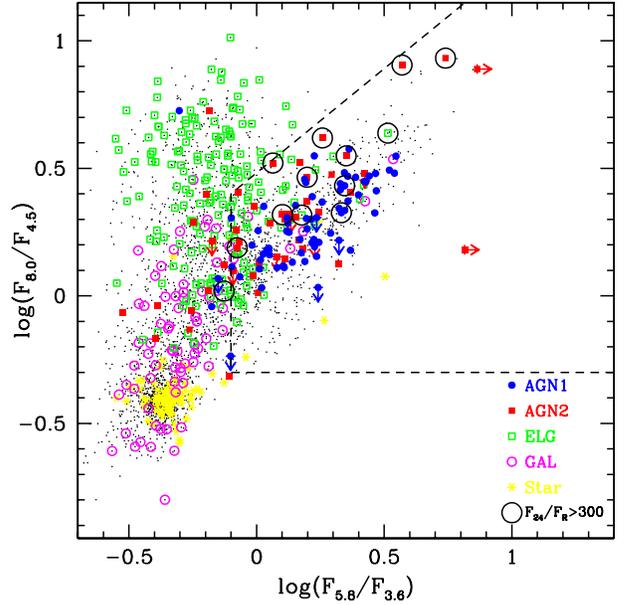}
     \caption{IRAC color--color diagram of the spectroscopically classified 3.6 $\mu$m sources.
     Black points show all the sources detected in all the four IRAC bands. The dashed line
     shows the AGN selection criteria proposed by Lacy et al. (2004, 2007). Open black circles
     indicate those 24$\mu$m sources having \emph{F}(24 $\mu$m)/\emph{F}(\emph{R}) ratios larger than 300.
     For more information about the sources distribution, see, e.g., Sajina et al. (2005).}
        \label{Fig_Lacy}
  \end{figure}

With the launch of {\em Spitzer}, some color--color diagrams
(mainly based on the IRAC photometry) have been proposed in order
to quickly characterize the MIR SED properties of the sources and
select the AGN (see, e.g.,  Lacy et al. 2004; Stern et al. 2005;
Lacy et al. 2007). In Figure \ref{Fig_Lacy}, the IRAC color--color
diagram, as proposed by Lacy et al. (2004, 2007), of all the
spectroscopically classified 3.6 $\mu$m sources is shown. The
dashed line shows the original AGN selection criteria, calibrated
on a Sloan Digital Sky Survey quasar sample and subsequently confirmed with
spectroscopic observations. The study of the relations among
the average SED properties of the sources and their optical
spectra is beyond the scope of this paper. However, we can note
that, as in the above mentioned studies, also the sources of our
sample occupy two branches, forming a ``V''-shaped locus. Each
spectroscopic class preferentially occupies different regions of
the diagram according to the energetically dominant component in
the MIR regime (see, e.g., Sajina et al. 2005). In line with
the above-mentioned studies, $\sim$85\% of the AGN (97\% of the
AGN1 and 70\% of the AGN2 detected in all the four IRAC bands) in
our sample are located inside the AGN selection region (the branch
on the right) corresponding to sources having red power-law
MIR-SEDs whose slope increases moving to the upper right part of
the diagram. ELG and GAL occupy (without solution of continuity)
the upper and lower region, respectively, of the left branch.
Stars are located at the bottom, almost in the center of the
passive galaxy locus, as they have qualitatively similar SEDs
(with an emission shortfall in the MIR).

%Fig 11

%%%%%%%%%%%%%%%%%%%%%%%%%%%%%%%%%%%%%%%%%%%%%%%
\section{Fraction of AGN}\label{cap:fractionAGN}

As discussed in the introduction, it is very interesting to measure the fraction of AGN among those
sources showing large \emph{F}(24 $\mu$m)/\emph{F}(\emph{R}) ratios or as a function of the 24 $\mu$m flux.
In order to use the most statistically useful sample, in the following analyses we will use the sub-sample of the spectroscopic catalog of the 24 $\mu$m extragalactic sources corresponding to the
intersection of the area of the spectroscopic follow up (see section \ref{par:24umsample}) with the area covered by the
\emph{XMM-Newton} observations (see also Figure \ref{Fig_radecAllz}).
As discussed in section \ref{par:24umsample}, we select a sub-sample at 24 $\mu$m fluxes
brighter than 280 $\mu$Jy and magnitudes brighter than \emph{R}=24.2,  which corresponds to the largest
possible sample where sufficient spectroscopic information exists.
These selections correspond
to an area of 0.54 deg$^2$.

\vspace{1.2cm}

%%%%%%%%%%%%%%%%%%%%%%%%%%%%%%%%%%%%%%%%%%%%%%%%%%%%
\subsection{Fraction of AGN as a function of the F(24~$\mu$m)/\emph{F}(\emph{R}) ratio}

The spectroscopic campaign has been carried out observing sources detected by
either \emph{XMM-Newton} or {\em Spitzer} (or both). 83  of the 419 spectroscopically identified 24 $\mu$m sources are also
included in the {\it XMM-Newton} detection catalog (see Table \ref{table_runs}).
Therefore, in the estimate of the AGN fraction as a function of the F(24~$\mu$m)/\emph{F}(\emph{R}) ratio,
a correction
for the bias introduced by the inclusion of the X-ray sample has to be  taken into account (indeed, as shown by Feruglio et al. (2008),
most of  the X-ray sources are AGN). If we call \emph{N}$_{{\rm X}}$ (and \emph{N}$_{{\rm X,spec}}$)  and \emph{N}$_{{\rm NoX}}$ (and \emph{N}$_{{\rm NoX,spec}}$) the number of {\em Spitzer} sources
(and those spectroscopically identified) in a given MIR flux interval, detected and not detected in the  X-ray band, respectively, the true fraction of AGN (\emph{Fr}$_{{\rm AGN}}$)
among all (\emph{N}$_{\rm Tot}$) the {\em Spitzer} sources is given by

$$ {Fr}_{\rm AGN} = \left( {{N}_{\rm X,AGN} \over {N}_{\rm X,spec}}   {{N}_{\rm X} } +   {{N}_{\rm NoX,AGN}\over{N}_{\rm NoX,spec}}    {{N}_{\rm NoX}} \right) / {N}_{\rm Tot}\ ,$$

\noindent
where \emph{N}$_{{\rm X,AGN}}$ and \emph{N}$_{{\rm NoX,AGN}}$ are the number of spectroscopically identified AGN, detected and not detected in the X-ray band, respectively.

\begin{figure}[]
  \centering
  \includegraphics[width=8.5cm, angle=0]{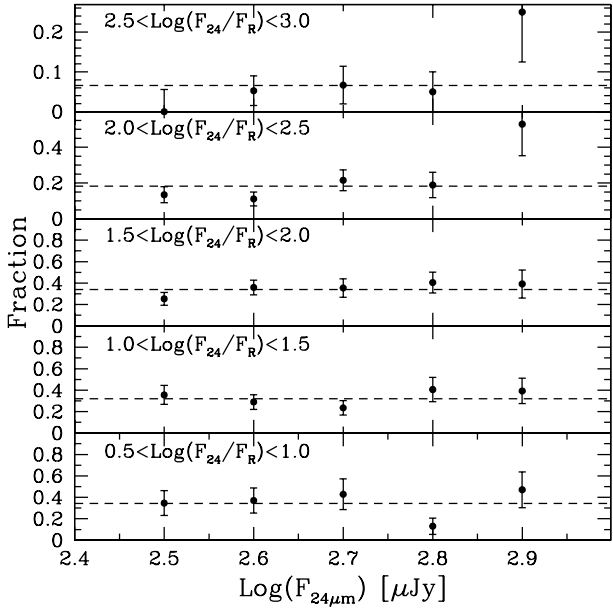}
     \caption{Fraction of spectroscopically identified 24 $\mu$m sources  as a function of the \emph{F}(24 $\mu$m)
flux divided into different bins of the  \emph{F}(24 $\mu$m)/\emph{F}(\emph{R}) ratio (the dashed horizontal lines indicate
mean values in each \emph{F}(24 $\mu$m)/\emph{F}(\emph{R}) bin).
Given the narrow dimension of the intervals in
\emph{F}(24 $\mu$m)/\emph{F}(\emph{R}), the overall constant trend in \emph{F}(24 $\mu$m) is similar in \emph{F}(\emph{R}) as well.
}
        \label{Fig_Frac_Spec}
  \end{figure}

We have verified that, although the fraction of spectroscopic identifications decreases at faint
\emph{R}-band magnitudes, in each \emph{F}(24 $\mu$m)/\emph{F}(\emph{R}) bin the fraction of spectroscopic identifications is not
significantly dependent on either the 24 $\mu$m flux or the \emph{R}-band magnitude (see Figure
\ref{Fig_Frac_Spec}). Therefore, the spectroscopic identifications provide a fair random sampling of the
sources in each \emph{F}(24 $\mu$m)/\emph{F}(\emph{R}) bin.

%Fig 12

%Fig 13
  \begin{figure}[]
  \centering
  \includegraphics[height=8.5cm, angle=-90]{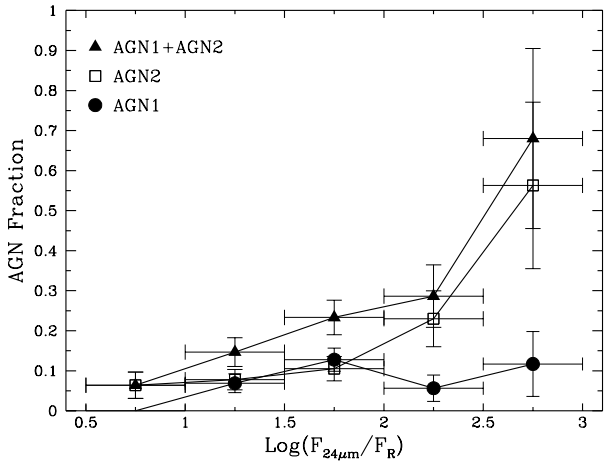}
     \caption{Fraction of AGN as a function of the \emph{F}(24 $\mu$m)/\emph{F}(\emph{R}) ratio (triangles).
     Filled circles represent the fraction of AGN1, while open squares represent the fraction
     of AGN2. Horizontal error bars represent the
     ratio interval; vertical bars represent the Poissonian errors
     propagated through the $Fr_{\rm AGN}$ formula in Section 5.1.}
        \label{Fig_fracAGN_24R}
  \end{figure}

\begin{figure}[]
  \centering
  \includegraphics[height=8.5cm,angle=-90]{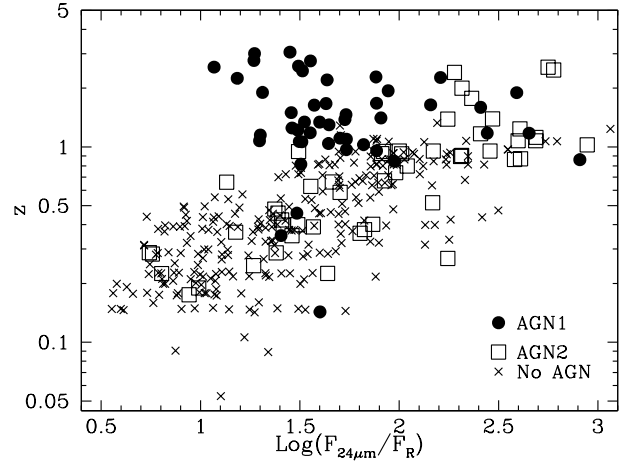}
     \caption{Redshift of the spectroscopically identified sources as a function of the \emph{F}(24 $\mu$m)/\emph{F}(\emph{R}) ratio.
    Filled circles and open squares represent AGN1 and AGN2, respectively. }
        \label{Fig_Z_24R}
  \end{figure}

The  fraction of AGN as a function of the \emph{F}(24 $\mu$m)/\emph{F}(\emph{R}) ratio
is shown in  Figure \ref{Fig_fracAGN_24R}, while in Figure
\ref{Fig_Z_24R} the redshift of the spectroscopically identified
sources as a function of the  \emph{F}(24 $\mu$m)/\emph{F}(\emph{R}) ratio is plotted.
The fraction of AGN clearly increases with increasing \emph{F}(24 $\mu$m)/\emph{F}(\emph{R}) ratio. While 85\% of the AGN1 concentrates in
the range 1$<$log[\emph{F}(24 $\mu$m)/\emph{F}(\emph{R})]$<$2, which is typical of the
local AGN1 SEDs (e.g., Spinoglio et al. 2002), the
AGN2 fraction constantly increases with the \emph{F}(24 $\mu$m)/\emph{F}(\emph{R})
ratio: in the 2.5$<$log[\emph{F}(24 $\mu$m)/\emph{F}(\emph{R})]$<$3 interval (where a
total of 18 sources have been spectroscopically identified)  AGN2
outnumber AGN1 by a factor of $\sim$5, and the total fraction of AGN
is 70($\pm$ 20)\%.

%Fig 14

However, it should be noted that this statistics could be
affected by some biases. In poor-quality data, emission line
spectra are more easily identified than no-emission line spectra.
In the 2.5$<$log[\emph{F}(24 $\mu$m)/\emph{F}(\emph{R})]$<$3 interval, beside the 18
classified spectra, 8 further FORS2 spectra  (with similar \emph{R}-band
distribution) missed an identification. These additional 8
spectra could be populated by the same mixture of AGN1, AGN2, ELG,
and passive galaxies, but, under the most extreme assumption that
all these objects were included among the non-AGN sample,
the fraction  of AGN in the 2.5$<$log[\emph{F}(24 $\mu$m)/\emph{F}(\emph{R})]$<$3
interval would be 50\%, instead of 70\%. Therefore, although
our estimate of a fraction of 70\% AGN in the 2.5$<$log[\emph{F}(24 $\mu$m)/\emph{F}(\emph{R})]$<$3 interval has not to be considered an upper
limit, it should be bear in mind that if this bias were present,
the real dependence of the AGN fraction with increasing \emph{F}(24 $\mu$m)/\emph{F}(\emph{R}) ratio would be weaker (but still present) than shown
in Figure  \ref{Fig_fracAGN_24R}.

On the other hand, in Section 3.5, we have stated that a
fraction (which is difficult to quantify) of the ELG classified
sources could instead reveal to be AGN2 if identified with higher
S/N optical spectra. Although we are not able to quantify this
bias, we can qualitatively expect it to be more relevant for the
optically fainter sources, whose \emph{F}(24 $\mu$m)/\emph{F}(\emph{R}) ratio is on
average larger. In this case, the increase of the AGN fraction
with increasing \emph{F}(24 $\mu$m)/\emph{F}(\emph{R}) ratio would be even more
relevant.

%Fig 15
  \begin{figure*}[]
  \centering
  \includegraphics[width=15.5cm]{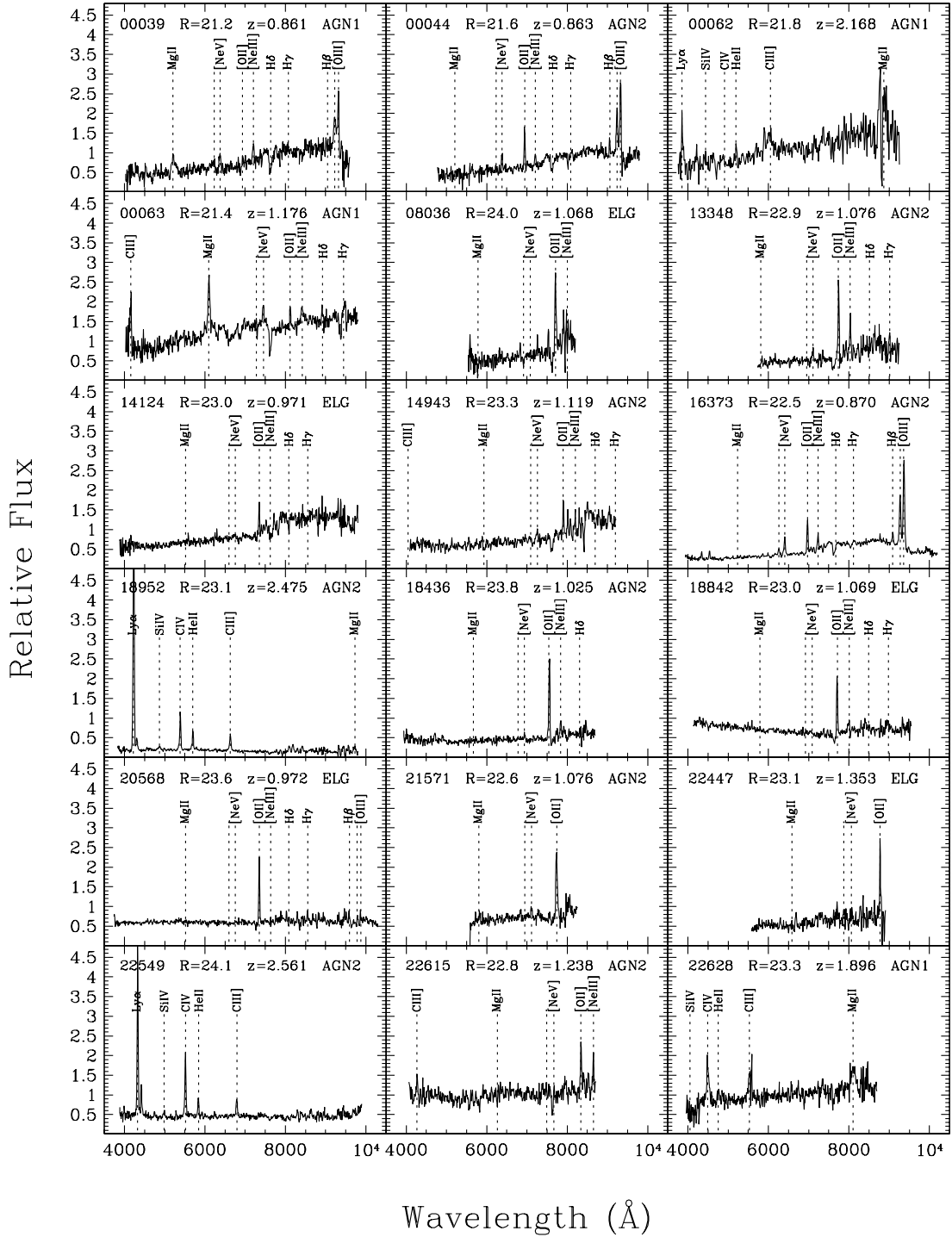}
     \caption{Optical spectra of the 18 sources with 2.5$<$log(\emph{F}(24 $\mu$m)/\emph{F}(\emph{R}))$<$3.  The positions of typical AGN emission lines
     (even if not observed) are shown as a reference.}
        \label{Fig_Sp}
  \end{figure*}

In Figure \ref{Fig_Sp}, the optical spectra of the 18 sources with 2.5$<$log[\emph{F}(24 $\mu$m)/\emph{F}(\emph{R})]$<$3
are shown. The positions of typical AGN emission lines (even if not observed) are shown as a reference.
Ten out of the 14 spectra with magnitudes larger than \emph{R}$=$22.5 mag have been observed by FORS2.
The AGN2 show larger \Oiii/\Hb\  and  \Oii/\Hb\ ratios than observed
in the ELG, typical of AGN activity (e.g., Osterbrock 1989).  Moreover, when observable, AGN2 show the high-ionization  \Nev\  emission line, which instead is not present in the ELG spectra (see also Figure \ref{Fig_CompoSp}).

The observed differences on the  \emph{F}(24 $\mu$m)/\emph{F}(\emph{R}) ratios of AGN1
and AGN2  is principally explained by the difference between their
optical luminosity. While in AGN1 the optical luminosity
originates directly in the nucleus, in AGN2,  as the AGN is
obscured, the optical luminosity comes from the lower luminous
hosting galaxy. As shown in Figures \ref {Fig_L24_z} and
\ref{Fig_Z_24R},  for non-AGN1 sources  the average redshift
increases with increasing \emph{F}(24 $\mu$m)/\emph{F}(\emph{R}) ratio. All the sources
with log[\emph{F}(24 $\mu$m)/\emph{F}(\emph{R})]$>$2.5 have \emph{z}$>$0.8, and $\nu$L$_\nu$
24 $\mu$m luminosity larger than 10$^{44}$ erg s$^{-1}$.  This
behavior is due to the fact that those non-AGN1 sources showing
larger \emph{F}(24 $\mu$m)/\emph{F}(\emph{R}) ratio have preferentially the faintest
optical magnitudes and thus the largest redshift. In contrast, AGN1 are more easily found at large redshifts but with
lower (roughly constant) \emph{F}(24 $\mu$m)/\emph{F}(\emph{R}) ratios. In this case
(as discussed in Section 4.1), this is caused by  the evolution of
AGN1 luminosity functions (the density increases with increasing
redshift\footnote{A luminosity evolution results in an increase of
the density of sources with the same luminosity.}) combined with
the accessible volumes and \emph{k}-corrections (see, e.g., Matute et al.
2002, 2006), while the average SED (and then the \emph{F}(24 $\mu$m)/\emph{F}(\emph{R})
ratio) does not evolve significantly.

\subsubsection{IRAC--MIPS color--color selection of absorbed AGN}

As shown in Figure \ref{Fig_Lacy}, the sources with large ($>$300) \emph{F}(24 $\mu$m)/\emph{F}(\emph{R}) ratios  are mostly located in the IRAC color--color
region where AGNs are expected: i.e., in the branch where both the \emph{F}(5.8 $\mu$m)/\emph{F}(3.6 $\mu$m) and  \emph{F}(8.0 $\mu$m)/\emph{F}(4.5 $\mu$m) ratios are larger.
However, it is evident that the \emph{F}(5.8 $\mu$m)/\emph{F}(3.6 $\mu$m) ratio is more efficient than the \emph{F}(8.0 $\mu$m)/\emph{F}(4.5 $\mu$m) ratio in the AGN selection.
It is therefore possible to create an optical--IRAC--MIPS color--color diagram,
based on the   \emph{F}(5.8 $\mu$m)/\emph{F}(3.6 $\mu$m) and \emph{F}(24 $\mu$m)/\emph{F}(\emph{R}) ratios, which is highly efficient in selecting obscured AGN (AGN2).

%Fig 16
 \begin{figure}[]
  \centering
  \includegraphics[width=8.5cm]{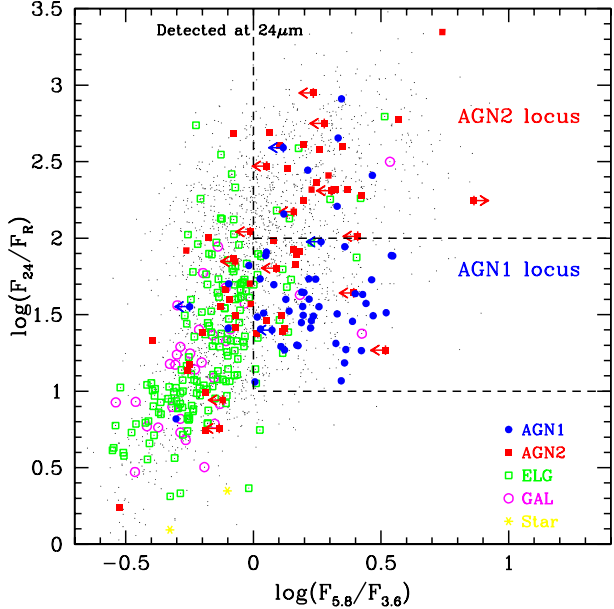}

     \caption{\emph{F}(24 $\mu$m)/\emph{F}(\emph{R}) ratio as a function of the (IRAC) \emph{F}(5.8 $\mu$m)/\emph{F}(3.6 $\mu$m) ratio
     of 24 $\mu$m detected sources.
     The loci mostly populated by AGN1 and obscured AGN2 are delimited by dashed lines
     (the respective fractions are reported in the text).
     The larger \emph{F}(24 $\mu$m)/\emph{F}(\emph{R}) ratio of AGN2 compared to AGN1 is due to the stronger obscuration
     in the optical for AGN2, while the \emph{F}(5.8 $\mu$m)/\emph{F}(3.6 $\mu$m) separation between ELG and AGN is
     mainly due to the PAH features present in most IR galaxies combined with a redshift effect (for more information, see, e.g., Sajina et al. 2005).
     The classification of the spectroscopically identified sources is shown in the
     legend.}
   \label{Fig_Nostro}
  \end{figure}

Figure \ref{Fig_Nostro} shows how this optical--IRAC--MIPS diagram
is populated according to our spectroscopic classification.
The region with log[\emph{F}(5.8 $\mu$m)/\emph{F}(3.6 $\mu$m)]$>$0 and
1$<$log[\emph{F}(24 $\mu$m)/\emph{F}(\emph{R})]$<$2 is mainly populated by AGN1: 65\%
are AGN1 and 12\% are AGN2 (79\% of all the AGN1 are located in
this region). The region with log[\emph{F}(5.8 $\mu$m)/\emph{F}(3.6 $\mu$m)]$>$0
and log[\emph{F}(24 $\mu$m)/\emph{F}(\emph{R})]$>$2 is mainly populated by AGN2: 24\%
are AGN1 and 55\% are AGN2 (36\% of all the AGN2 are included in
this area). Therefore, while the first region is quite efficient
in selecting AGN1, the second one is useful in selecting obscured
AGN2 with a moderate level of completeness.

%%%%%%%%%%%%%%%%%%%%%%%%%%%%%%%%%%%%%%%%%%%%%%%
\subsection{Fraction of AGN as a function of the 24 $\mu$m flux}

The measure of the fraction (and counts) of AGN as a function of the 24 $\mu$m flux is a very useful information
to constrain the evolutionary models  of AGN and starburst galaxies (see, e.g., Gruppioni et al. 2005; Franceschini et al. 2008).
Unlike for the estimate of the AGN fraction as a function of the  \emph{F}(24 $\mu$m)/\emph{F}(\emph{R}) ratio (presented in the previous section), in this case, the presence in our sample of a decrease of the fraction of spectroscopically
identified sources with decreasing \emph{R}-band magnitudes (see Figure \ref{Fig_RvsF24}) and, on top of that, the lack of any identification
at magnitudes fainter than \emph{R}=24.2, complicates the attainment of a reliable measure.

In order to overcome these difficulties, and still obtain a sufficiently useful estimate, we have derived
the fraction of AGN as a function of the 24 $\mu$m flux by assuming that it
is mainly dependent on the \emph{F}(24 $\mu$m)/\emph{F}(\emph{R}) ratio as measured in the previous section.
The AGN fraction was then derived by dividing each 24 $\mu$m flux interval into
\emph{F}(24 $\mu$m)/\emph{F}(\emph{R}) bins and multiplying the number of sources in each bin to its previously
measured  fraction of AGN (see Figure \ref{Fig_fracAGN_24R}).

%Fig 17
  \begin{figure}[]
  \centering
  \includegraphics[height=8.5cm, angle=-90]{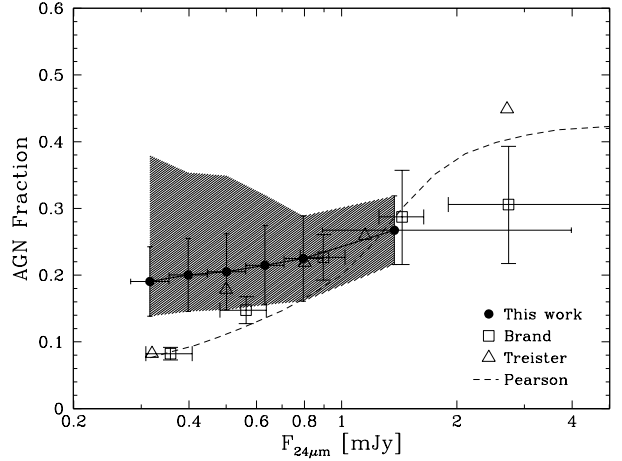}
     \caption{Fraction of optically classified AGN as a function of the 24 $\mu$m flux. The shaded area shows the 1$\sigma$ confidence limits  once the sources not detected in the \emph{R}-band are included and assumed to contain a fraction of
68\% of AGN, as measured in the  2.5$<$log[\emph{F}(24 $\mu$m)/\emph{F}(\emph{R})]$<$3 bin.
Overplotted are the estimates based on both X-ray and {\em Spitzer} data from Brand et al. (2006; squares) and Treister et al. (2006; triangles).  The dashed line shows the expectations from the burst model of Pearson (2005).  Horizontal error bars of our data represent the flux interval.
     }
        \label{Fig_fracAGN_24}
  \end{figure}

In Figure \ref{Fig_fracAGN_24}, the corresponding fraction of optically classified AGN as a function of the
 24 $\mu$m flux is shown.  The shaded area corresponds to the 1 $\sigma$  confidence limits
     once the sources not detected in the \emph{R}-band are also included and assumed to contain a fraction of
68\% of AGN, as measured for the objects with the largest \emph{F}(24 $\mu$m)/\emph{F}(\emph{R}) ratio. The AGN fraction at \emph{F}(24 $\mu$m)$\sim$0.8
mJy results to be $\sim$22($\pm$7)\%  and decreases slowly to
$\sim$19($\pm$5)\% down to \emph{F}(24 $\mu$m)$\sim$0.3 mJy. This
estimate of the  fraction of optically classified AGN, at 24
$\mu$m fluxes fainter than 0.8 mJy, is larger than the previous
measures by Brand et al. (2006) and Treister et al. (2006) and
the expectations of the burst model of Pearson (2005) (see Figure
\ref{Fig_fracAGN_24}). It should be noted that our measure of
the AGN fraction is based on a direct spectroscopic optical
classification, while the estimates from  Brand et al. (2006)
are based on a classification which uses {\em Spitzer} colors
(selecting sources in which the MIR emission is AGN dominated),
and Treister et al. (2006) use a selection in the hard-X band
and then corrects by the absorbed AGN fraction expected to be
missed assuming an $N_H$ population distribution. The real
fraction of AGN would be even larger if we took into account that
a fraction ($\sim$10\%) of X-ray (or even MIR) classified AGN do
not show in the optical signatures of AGN activity (e.g., the
XBONG; see Fiore et al. 2000, 2003; La Franca et al. 2007; Cocchia
et al. 2007; Caccianiga et al. 2008; Gruppioni et al. 2008;
Feruglio et al. 2008 and references therein). These objects
usually show large values of both X-ray to optical and MIR to
optical flux ratios.

%%%%%%%%%%%%%%%%%%%%%%%%%%%%%%%%%%%%%%%%%%%%%%%
\section{Discussion and Conclusions}

As discussed in the introduction, several studies have presented
evidence that many galaxies showing SEDs  with large \emph{F}(24 $\mu$m)/\emph{F}(\emph{R}) ratios (e.g., Polletta et al. 2008 use a threshold of
\emph{F}(24 $\mu$m)/\emph{F}(\emph{R})$>$400 and 24 $\mu$m fluxes larger than 1
mJy) harbor an AGN. These works were mainly based on low-resolution MIR spectroscopy or SED studies. X-ray stacking
analyses of sources with \emph{F}(24 $\mu$m)/\emph{F}(\emph{R})$>$1000 have shown that
their {\em average}  X-ray spectrum is compatible with
Compton-thin and Compton-thick absorbed sources (Daddi et al. 2007; Fiore
et al. 2008, 2009), but the relevance of this result depends on
the fraction of AGN (and complementary  starburst galaxies) which
contribute to the average X-ray spectrum (see Donley et al. 2008).

Our catalog of 1376 optical spectroscopic identifications of 3.6
and 24 $\mu$m SWIRE sources, thanks to the use of  FORS2 and VIMOS
at VLT, has pushed the optical classification down to $R\sim24$,
and has allowed us to  directly  measure the fraction of AGN as a
function of the \emph{F}(24 $\mu$m)/\emph{F}(\emph{R}) ratio. We have then shown that,
at 24 $\mu$m fluxes larger than 280 $\mu$Jy, in the range
316$<$\emph{F}(24 $\mu$m)/\emph{F}(\emph{R})$<$1000, 70($\pm$20)\% of the sources show
an optical AGN spectrum, and most of them are AGN2. In fact, the
increase of the total fraction of AGN with the \emph{F}(24 $\mu$m)/\emph{F}(\emph{R})
ratio is caused by a strong increase of the fraction of AGN2,
which populate more than 80\% of the AGN found in the
316$<$\emph{F}(24 $\mu$m)/\emph{F}(\emph{R})$<$1000 range. This result is in agreement
with the above-mentioned observed average X-ray absorbed spectrum
reported for similar kind of sources. Indeed, in the framework of
the classical unified AGN scenario X-ray absorbed spectra are
typical of AGN2.

At fainter optical luminosity (\emph{R}$>$24) and then larger \emph{F}(24 $\mu$m)/\emph{F}(\emph{R})
ratios the fraction of AGN might be even higher,  but the optical spectroscopic identification
of a statistical significant sample of these sources is difficult to achieve with the currently available 8~m class telescopes.

Our result confirms previous indications that the population of
sources showing  large \emph{F}(24 $\mu$m)/\emph{F}(\emph{R}) ratios are mainly AGN2
(e.g., Daddi et al. 2007; Fiore et al. 2008, 2009), but as we
were able to spectroscopically separate the AGN(2) from the
starburst galaxies, we are now able to measure how many and how
much of these AGNs show X-ray absorption and, even more
interestingly, if they include a significant fraction of the  long
searched and predicted Compton-thick AGN population. Such a  kind
of study requires the use of deep {\em Chandra} X-ray observations
which, in ELAIS-S1, partly already exist (S. Puccetti et al., in
preparation), and will be the subject of a  forthcoming paper.

\vspace{0.4cm}

\noindent {\it Acknowledgments:}
Based on observations made with the ESO telescopes at the La Silla
and Paranal Observatories under program IDs 168.A-0322, 170.A-0143,
073.A-0446, 075.A-0428, 076.A-0225, 077.A-0800, and 078.A-0795.
Part of the data published in this paper have been reduced using VIPGI,
designed by the VIRMOS Consortium and developed by INAF Milano.
We are grateful to Bianca Garilli and Marco Scodeggio for the support
provided in running the VIPGI pipeline.
This work is based on observations made with the \emph{Spitzer Space Telescope},
which is operated by the Jet Propulsion Laboratory, California Institute
of Technology under a contract with NASA. We acknowledge
financial contribution from contract ASI-INAF I/023/05/0, PRIN-MIUR grant 2006-02-5203.
The anonymous referee is acknowledged for his detailed and constructive
reports.

\vfill

\noindent
{\it Facilities:} \facility{VLT:Melipal (VIMOS)},
\facility{VLT:Antu (FORS2)}, \facility{ESO:3.6\ m (EFOSC2)},
\facility{\emph{Spitzer}}

%\vfill\eject


\begin{thebibliography}{}

\bibitem[Alexander et al.(2001)]{2001ApJ...554...18A} Alexander, D.~M., et
al.\ 2001, \apj, 554, 18

\bibitem[Alexander et al.(2008)]{2008ApJ...687..835A} Alexander, D.~M., et
al.\ 2008, \apj, 687, 835

\bibitem[Alonso-Herrero et al.(2006)]{2006ApJ...640..167A} Alonso-Herrero,
A., et al.\ 2006, \apj, 640, 167

\bibitem[Ballantyne et al.(2006)]{2006ApJ...653.1070B} Ballantyne, D.~R.,
Shi, Y., Rieke, G.~H., Donley, J.~L., Papovich, C.,
\& Rigby, J.~R.\ 2006, \apj, 653, 1070

\bibitem[Berta et al. (2006)]{Bert06} Berta, S., et al. 2006, \aap, 451, 881

\bibitem[Berta et al. (2008)]{Bert08} Berta, S., et al. 2008, \aap, 488, 533

\bibitem[Brand et al.(2006)]{2006ApJ...644..143B} Brand, K., et al.\ 2006,
\apj, 644, 143


\bibitem[Brand et al.(2007)]{2007ApJ...663..204B} Brand, K., et al.\ 2007,
\apj, 663, 204

\bibitem[Brand et al.(2008)]{2008ApJ...680..119B} Brand, K., et al.\ 2008,
\apj, 680, 119

\bibitem[Brown et al.(2006)]{2006ApJ...638...88B} Brown, M.~J.~I., et al.\
2006, \apj, 638, 88

\bibitem[Burgarella et al.(2005)]{2005ApJ...619L..63B} Burgarella, D., et
al.\ 2005, \apjl, 619, L63

\bibitem[Caccianiga et al.(2008)]{2008A&A...477..735C} Caccianiga, A., et al.\ 2008,
\aap, 477, 735

\bibitem[Caputi et al.(2006)]{2006ApJ...637..727C} Caputi, K.~I., et al.\
2006, \apj, 637, 727

\bibitem[Cocchia et al.(2007)]{2007A&A...466...31C} Cocchia, F., et al.\ 2007, \aap,
466, 31

\bibitem[Croom et al.(2004)]{2004MNRAS.349.1397C} Croom, S.~M., Smith,
R.~J., Boyle, B.~J., Shanks, T., Miller, L., Outram, P.~J.,
\& Loaring, N.~S.\ 2004, \mnras, 349, 1397

\bibitem[Daddi et al. (2007)]{Dad07} Daddi, E., et al. 2007, \apj, 670, 173

\bibitem[Davis et al.(2007)]{2007ApJ...660L...1D} Davis, M., et al.\ 2007,
\apjl, 660, L1

\bibitem[Dey et al.(2008)]{2008ApJ...677..943D} Dey, A., et al.\ 2008,
\apj, 677, 943

\bibitem[Desai et al.(2008)]{2008ApJ...679.1204D} Desai, V., et al.\ 2008,
\apj, 679, 1204

\bibitem[Donley et al.(2005)]{2005AJ....129..220D} Donley, J.~L., et al.\
2005, \aj, 129, 220


\bibitem[Donley et al.(2007)]{2007ApJ...660..167D} Donley, J.~L., Rieke,
G.~H., P{\'e}rez-Gonz{\'a}lez, P.~G., Rigby, J.~R.,
\& Alonso-Herrero, A.\ 2007, \apj, 660, 167

\bibitem[Donley et al. (2008)]{2008} Donley, J.~L., Rieke,
G.~H., P{\'e}rez-Gonz{\'a}lez, P.~G., Barro \ 2008, \apj,  687, 111

\bibitem[Fadda et al.(2006)]{2006AJ....131.2859F} Fadda, D., et al.\ 2006,
\aj, 131, 2859

\bibitem[Fazio et al.(2004)]{2004ApJS..154...10F} Fazio, G.~G., et al.\
2004, \apjs, 154, 10

\bibitem[Feruglio et al. (2008)]{Feruglio2008} Feruglio, C. et al. 2008, \aap, 488, 417

\bibitem[Fiore et al.(2000)]{2000NewA....5..143F} Fiore, F., et al.\ 2000,
New Astron., 5, 143

\bibitem[Fiore et al. (2003)]{fiore03} Fiore F., et al. 2003, \aap, 409, 79

\bibitem[Fiore et al. (2008)]{fiore08} Fiore F., et al. 2008, \apj, 672, 94


\bibitem[Fiore et al. (2009)]{fiore09} Fiore F., et al. 2009, \apj, 693, 447


\bibitem[Franceschini et al.(2005)]{2005AJ....129.2074F} Franceschini, A.,
et al.\ 2005, \aj, 129, 2074

\bibitem[Franceschini et al.(2008)]{} Franceschini, A.,
et al.\ 2008, \aap, 487, 837

\bibitem[Gruppioni et al.(1999)]{1999MNRAS.305..297G} Gruppioni, C., et
al.\ 1999, \mnras, 305, 297

\bibitem[Gruppioni et al.(2005)]{2005ApJ...618L...9G} Gruppioni, C., Pozzi, F., Lari, C., Oliver, S., \& Rodighiero, G.\ 2005, \apjl, 618, L9

\bibitem[Gruppioni et al.(2008)]{Gru2008} Gruppioni, C. et al.\ 2008, \apj, 684, 136

\bibitem[Hasinger(2008)]{2008A&A...490..905H} Hasinger, G.\ 2008, \aap, 490, 905

\bibitem[Hatziminaoglou et al.(2005)]{2005AJ....129.1198H} Hatziminaoglou, E., et al.\ 2005, \aj, 129, 1198

\bibitem[Houck et al.(2005)]{2005ApJ...622L.105H} Houck, J.~R., et al.\
2005, \apjl, 622, L105

\bibitem[Kewley, L. J. et al.(2006)]{2006MNRAS.372..961K} Kewley, L.~J. et al.\ 2006, \mnras, 372, 961

\bibitem[La Franca et al.(2004)]{2004AJ....127.3075L} La Franca, F., et
al.\ 2004, \aj, 127, 3075

\bibitem[La Franca et al.(2005)]{2005ApJ...635..864L} La Franca, F., et
al.\ 2005, \apj, 635, 864

 \bibitem[La Franca et
al.(2007)]{2007A&A...472..797L} La Franca, F., et al.\ 2007, \aap, 472, 797

\bibitem[Lacy et al.(2004)]{2004ApJS..154..166L} Lacy, M., et al.\ 2004,
\apjs, 154, 166

\bibitem[Lacy et al.(2007)]{2007AJ....133..186L} Lacy, M., Petric, A.~O.,
Sajina, A., Canalizo, G., Storrie-Lombardi, L.~J., Armus, L., Fadda, D.,
\& Marleau, F.~R.\ 2007, \aj, 133, 186

\bibitem[Lawrence
\& Elvis(1982)]{1982ApJ...256..410L} Lawrence, A., \& Elvis, M.\ 1982, \apj, 256, 410

\bibitem[2001]{lari01} Lari, C., et al. 2001, MNRAS, 325, 1173

\bibitem[Lonsdale et al.(2003)]{2003PASP..115..897L} Lonsdale, C.~J., et
al.\ 2003, \pasp, 115, 897

\bibitem[Lonsdale et al.(2004)]{Lons04} Lonsdale, C.J. et al. 2004, \apjs, 154, 54

\bibitem[Mart{\'{\i}}nez-Sansigre et al.(2005)]{2005Natur.436..666M}
Mart{\'{\i}}nez-Sansigre, A., Rawlings, S., Lacy, M., Fadda, D., Marleau,
F.~R., Simpson, C., Willott, C.~J., \& Jarvis, M.~J.\ 2005, \nat, 436, 666

\bibitem[Mart{\'{\i}}nez-Sansigre et al.(2007)]{2007MNRAS.379L...6M}
Mart{\'{\i}}nez-Sansigre, A., et al.\ 2007, \mnras, 379, L6

\bibitem[Matute et al.(2002)]{2002MNRAS.332L..11M} Matute, I., et al.\
2002, \mnras, 332, L11


\bibitem[Matute et al.(2006)]{2006A&A...451..443M} Matute, I., La Franca, F.,
Pozzi, F., Gruppioni, C., Lari, C., \& Zamorani, G.\ 2006, \aap, 451, 443


\bibitem[Middelberg et al.(2008)]{2008AJ....135.1276M} Middelberg, E., et
al.\ 2008, \aj, 135, 1276

\bibitem[Murray et al.(2005)]{2005ApJS..161....1M} Murray, S.~S., et al.\
2005, \apjs, 161, 1

\bibitem[Oliver et al.(2000)]{2000MNRAS.316..749O} Oliver, S., et al.\
2000, \mnras, 316, 749

\bibitem[Osterbrock (1989)]{osterbrock89} Osterbrock, D.\,E. 1989, Astrophysics of Gaseous Nebulae and Active
                        Galactic Nuclei (Mill Valley, CA: Univ. Science Books)

\bibitem[Pearson(2005)]{2005MNRAS.358.1417P} Pearson, C.\ 2005, \mnras,
358, 1417

\bibitem[P{\'e}rez-Gonz{\'a}lez et al.(2005)]{2005ApJ...630...82P}
P{\'e}rez-Gonz{\'a}lez, P.~G., et al.\ 2005, \apj, 630, 82

\bibitem[Polletta et al.(2006)]{2006ApJ...642..673P} Polletta, M., et
al.\ 2006, \apj, 642, 673

\bibitem[Polletta et al.(2007)]{2007ApJ...663...81P} Polletta, M., et al.\
2007, \apj, 663, 81

\bibitem[Polletta et al.(2008)]{2008ApJ...675..960P} Polletta, M., Weedman,
D., H{\"o}nig, S., Lonsdale, C.~J., Smith, H.~E.,
\& Houck, J.\ 2008, \apj, 675, 960


\bibitem[Pope et al.(2008)]{2008ApJ...689..127P} Pope, A., et al.\ 2008, \apj, 689, 127

\bibitem[Pozzi et al. (2004)]{Pozzi04} Pozzi, A. et al.  2004, \apj, 609, 122

\bibitem[Puccetti et al. (2006)]{Puc06} Puccetti, S. et al.  2006, \aap, 457, 501


\bibitem[Rieke et al.(2004)]{2004ApJS..154...25R} Rieke, G.~H., et al.\
2004, \apjs, 154, 25

\bibitem[Rowan-Robinson et al.(2004)]{2004MNRAS.351.1290R} Rowan-Robinson,
M., et al.\ 2004, \mnras, 351, 1290

\bibitem[Sajina et al.(2005)]{2005Apj..621.256S} Sajina, A., et
al.\ 2005, \apj, 621, 256

\bibitem[Sajina et al.(2006)]{2006S} Sajina, A., et
al.\ 2006, \mnras, 369, 939

\bibitem[Scodeggio et al.(2005)]{2005PASP..117.1284S} Scodeggio, M., et
al.\ 2005, \pasp, 117, 1284

\bibitem[Spinoglio et al.(2002)]{2002ApJ...572..105S} Spinoglio, L.,
Andreani, P., \& Malkan, M.~A.\ 2002, \apj, 572, 105

\bibitem[Steffen et al.(2007)]{2007ApJ...667L..25S} Steffen, A.~T., Brandt,
W.~N., Alexander, D.~M., Gallagher, S.~C.,
\& Lehmer, B.~D.\ 2007, \apjl, 667, L25


\bibitem[Stern et al.(2005)]{2005ApJ...631..163S} Stern, D., et al.\ 2005,
\apj, 631, 163

\bibitem[Surace et al.(2005)]{2005AAS...207.6301S} Surace, J.~A., Shupe,
D.~L., Fang, F., Evans, T., Alexov, A., Frayer, D., Lonsdale, C.~J.,
\& SWIRE Team 2005, BAAS, 37, 1246

\bibitem[Tajer et al.(2007)]{2007A&A...467...73T} Tajer, M., et al.\ 2007, \aap, 467, 73

\bibitem[Treister \& Urry(2006)]{2006ApJ...652L..79T} Treister, E., \& Urry, C.~M.\ 2006, \apjl, 652, L79

\bibitem[Treister et al.(2006)]{2006ApJ...640..603T} Treister, E., et al.\ 2006, \apj, 640, 603

\bibitem[Treister et al.(2009)]{} Treister, E., et al. 2009, \apj, 693, 1713

\bibitem[Tresse et al. (1996)]{Tresse96} Tresse, L., et al. 1996, MNRAS, 281, 847

\bibitem[Ueda et al. (2003)]{Ueda03} Ueda, Y., Akiyama, M.,
Ohta, K., \& Miyaji, T.\ 2003, \apj, 598, 886

\bibitem[Veilleux \& Osterbrock(1987)]{Veilleux87} Veilleux, S. \& Osterbrock, D.E. 1987, ApJS, 63, 295

 \bibitem[Weedman et al.(2006)]{2006ApJ...651..101W} Weedman, D.~W., et al.\
2006, \apj, 651, 101

\bibitem[Yan et al.(2007)]{2007ApJ...658..778Y} Yan, L., et al.\ 2007,
\apj, 658, 778


\end{thebibliography}
\end{document}